\documentclass[12pt,a4paper]{iopart}

\usepackage{graphicx, subcaption}
\usepackage{epstopdf}
\usepackage[utf8]{inputenc}
\usepackage[T1]{fontenc}
\expandafter\let\csname equation*\endcsname\relax
\expandafter\let\csname endequation*\endcsname\relax
\usepackage{amsmath, amsfonts,amssymb}
\usepackage{cancel,braket}
\usepackage{float}
\usepackage{slashed}
\usepackage{bbold}
\usepackage{hyperref}
\usepackage{tcolorbox} 
\usepackage{enumitem} 
\usepackage{schemata}
\usepackage{enumerate}
\usepackage{multirow}
\usepackage{pdfpages}
\usepackage{longtable}
\usepackage{parcolumns}
\usepackage{physics}
\usepackage{booktabs}
\usepackage{xcolor}
\usepackage{xfrac}
\usepackage{blindtext}
\usepackage{multicol}
\usepackage[toc,page]{appendix}

\newcommand{\blue}[1]{\textcolor{black}{#1}}
\newcommand{\red}[1]{\textcolor{black}{#1}}
\newcommand{\green}[1]{\textcolor{black}{#1}}

\begin{document}

\title[Comparison of gyrokinetic simulations with DR measurements in W7-X]{Electrostatic microturbulence in W7-X: comparison of local gyrokinetic simulations with Doppler reflectometry measurements}

\author[A. González-Jerez et al.]{A. González-Jerez$^1$, J. M. García-Regaña$^1$, I. Calvo$^1$, \\D. Carralero$^1$, T. Estrada$^1$, E. Sánchez$^1$, M. Barnes$^2$ and the W7-X Team}

\address{$^1$Laboratorio Nacional de Fusión, CIEMAT, 28040 Madrid, Spain}
\address{$^2$Rudolf Peierls Centre for Theoretical Physics, University of Oxford, Oxford OX1 3PU, UK}




\begin{abstract}
The first experimental campaigns of Wendelstein 7-X (W7-X) have shown that turbulence plays a decisive role in the performance of neoclassically optimized stellarators. This stresses the importance of understanding microturbulence from the theoretical and experimental points of view. To this end, this paper addresses a comprehensive characterization of the turbulent fluctuations by means of nonlinear gyrokinetic simulations performed with the code \texttt{stella} in two W7-X scenarios. In the first part of the paper, the amplitude of the density fluctuations is calculated and compared with measurements obtained by Doppler reflectometry (DR) in the OP1 experimental campaigns. It is found that the trend of the fluctuations along the radius is explained by the access of the DR system to different regions of the turbulence wavenumber spectrum. In the second part of the article, frequency spectra of the density fluctuations 
and the zonal component of the turbulent flow are numerically characterized for comparisons against future experimental analyses. Both quantities feature broad frequency spectra with dominant frequencies of O(1)-O(10) kHz.
   
\end{abstract}

\noindent{\it Keywords}: stellarator, gyrokinetic simulations, plasma turbulence, \\ Doppler reflectometry

\submitto{Nuclear Fusion}

\maketitle

\section{Introduction}
    Neoclassical transport has traditionally been a major source of concern with regard to energy losses, particle transport and impurity accumulation at the core of stellarators. In generic stellarators, trapped particles experience large radial excursions at low collisionality. In addition, ambipolarity is preserved by the onset of a radial electric field that, in relevant scenarios, leads to a strong inward convection of impurities. As a consequence, turbulent transport, driven by fluctuations with frequency much lower than the gyrofrequency of the plasma species and length scale of the order of the Larmor radius, has been explored in less detail in stellarators than in tokamaks. Nevertheless, the first experimental campaigns of the stellarator Wendelstein 7-X (W7-X) \cite{wolf2017, Klinger2019} ---the first large stellarator optimized for neoclassical transport--- have demonstrated that turbulence is an essential ingredient to understand how plasmas perform in this device. It has been observed that, in standard plasmas heated via electron-cyclotron-resonance-heating (ECRH), turbulent energy losses exceed by an order of magnitude the neoclassical ones \cite{Bozhenkov2020}. In addition, turbulent diffusion has been shown to underly the practical absence of strong accumulation of impurities, which, in general, exhibit a low peaking factor \cite{GarcaRegaa2021, Geiger2019}. The shape of the density profiles, with no sign of hollowness predicted by neoclassical theory \cite{Maasberg1999}, is explained by a robust turbulent pinch that manifests in broad regions of the parameter space \cite{Thienpondt2023}. In parallel to these observations and partly as a result of them, the interest in modelling gyrokinetic turbulence in stellarators has increased over the recent years. New codes have been developed \cite{Barnes2019, Maurer2020, Mandell2023}, verified against each other \cite{GonzlezJerez2022, Sanchez2021, Sanchez2023} and applied to fundamental questions in W7-X and other stellarators \cite{GarcaRegaa2021NF, ThienpondtISHW2022}. Nonetheless, comparisons of numerical gyrokinetic simulations against diagnostic measurements of turbulent fluctuations are scarce in stellarators and, in particular, in W7-X \cite{Bhner2021, Hansen2022}. \\

In this study, we present a numerical characterization of the turbulent fluctuations in W7-X, comparing them with the existing experimental measurements \cite{Carralero2021} obtained with the Doppler reflectometer (DR) system \cite{Windisch2018, Carralero2021, Estrada2021, Carralero2022}. This diagnostic is capable of providing the amplitude of density fluctuations ($|\delta n|$) and rotation velocity ($u_{\perp}$) at a specific spatial location in the plasma for a selected perpendicular wavenumber ($k_{\perp}^{\mathrm{DR}}$). In this work we use the gyrokinetic code \texttt{stella} \cite{Barnes2019} to simulate, in the radial range of measurement of the DR system, two different ECRH discharges by means of flux tube nonlinear simulations, considering both ions and electrons as kinetic species. The amplitude of the density fluctuations computed with \texttt{stella} is compared against measurements by the DR system. {In addition, we provide other quantities for future comparisons with the same diagnostic, in particular the zonal component of the turbulent $E\times B$ flow.} \\

The rest of the paper is organized as follows. In section \ref{sec:stella}, the features of the gyrokinetic code \texttt{stella} are briefly summarized. The coordinates used by the code are introduced and the equivalence between the wavenumber of measurement of the DR system and that used in \texttt{stella} is presented. In section \ref{sec:parameters}, the plasma profiles of the simulated ECRH discharges are described, as well as the configuration and parameters used in the simulations. In section \ref{sec:comparisons}, the squared amplitude of density fluctuations computed with \texttt{stella} for five different radial positions is compared against the measurements of the DR system analyzed in \cite{Carralero2021} for the two discharges studied. In section \ref{sec:predictions}, quantities suitable for future comparisons with the DR system are predicted. In particular, in subsection \ref{subsec:spectra_density}, the frequency spectra of density fluctuations expected to be measured by the DR system are given. In subsection \ref{subsec:spectra_ExB}, the frequency spectra of zonal flow fluctuations are obtained. Apart from aspects related to DR measurements, in section \ref{sec:potential}, linear and nonlinear simulations are presented in order to provide a picture of the turbulence in the two scenarios considered throughout this paper. Finally, section \ref{sec:conclusions} contains the summary and conclusions of this work.

\section{The code stella, coordinate system and correspondence with $k_{\perp}^{DR}$}\label{sec:stella}
    The gyrokinetic formalism \cite{Catto1978} is the theoretical framework to study microturbulence in strongly magnetized plasmas. It is based on the average over the gyrophase, removing the fastest degree of freedom of the system and retaining the effect of the finite Larmor radius size, which is comparable to the wavelength of turbulence. In this work we use the flux tube version of the gyrokinetic code \texttt{stella}. The flux tube domain fully exploits the scale separation between the typical variation lengths of the background magnetic field, plasma profiles and turbulent fluctuations along the magnetic field line with respect to the typical scales of the turbulent fluctuations in the plane perpendicular to the magnetic field. The code solves the gyrokinetic Vlasov and quasineutrality equations \cite{Barnes2019, GonzlezJerez2022} for an arbitrary number of species using a Fourier spectral treatment in the plane perpendicular to the magnetic field and a mixed implicit-explicit method for the calculation of the parallel streaming and acceleration terms in the gyrokinetic Vlasov equation. This mixed algorithm allows to use larger time steps than explicit methods, being this especially beneficial when kinetic electrons are considered. Usually, standard twist-and-shift boundary conditions \cite{Beer1995} are defined in the direction parallel to the magnetic field in flux tube codes. When the global magnetic shear of the device is small, these boundary conditions impose severe restrictions on the minimum wavelength considered in the direction perpendicular to the magnetic field or, equivalently, on the size of the flux tube in that direction. In particular, in W7-X (the device considered in this work), as one moves towards the magnetic axis, the global shear becomes critically low \blue{as it is shown in section \ref{sec:parameters}}. To avoid these restrictions and have control on the size of the flux tube, the stellarator symmetric twist-and-shift boundary conditions introduced in \cite{Martin2018} have been implemented in \texttt{stella} and applied to the analyses presented below.\\ 

The coordinates used in the code are denoted by $\{x,y,\zeta,v_{\|}, \mu\}$. We assume stellarator configurations with nested magnetic surfaces and define general spatial coordinates $\{r, \alpha, \zeta\}$, where $r\in [0,a]$ is a radial coordinate that labels magnetic surfaces, $\alpha \in [0,2\pi)$ is an angular coordinate labeling magnetic field lines on each surface and $\zeta \in [\zeta_{\mathrm{min}},\zeta_{\mathrm{max}}]$ is a coordinate along magnetic field lines. Here, $a$ is the minor radius of the device 
and $\zeta_{\min}$ and $\zeta_{\max}$ are the ends of the flux tube in the $\zeta$ coordinate. In particular, for $r$ we choose
\begin{equation}
    r:= a\sqrt{\frac{\psi_t}{\psi_{t,a}}}, 
\end{equation}
where $2\pi \psi_t(r)$ is the toroidal flux enclosed by the surface labeled by $r$ and $\psi_{t,a}:=\psi_t(a)$. For the coordinate $\alpha$, we take
\begin{equation}
    \alpha=\theta-\iota\zeta,
\end{equation}
where $\theta$ and $\zeta$ are, respectively, the poloidal and toroidal PEST flux coordinates \cite{Grimm1983} and $\iota (r)$ is the rotational transform. In these coordinates, the magnetic field takes the form
\begin{equation}
    \mathbf{B}=\frac{\dd \psi_t}{\dd r}\nabla r\times \nabla \alpha.
\end{equation}
If $r_0$ and $\alpha_0$ are the values of $r$ and $\alpha$ that define the flux tube (and, actually, the center of the simulation domain), the spatial coordinates $\{x, y\}$ in the plane perpendicular to $\mathbf{B}$ can be expressed as
\begin{equation}\label{x}
    x: = r - r_0
\end{equation}
and
\begin{equation}\label{y}
    y: = r_0(\alpha - \alpha_0).
\end{equation} 
Therefore, the modes $k_x$ and $k_y$ used in the Fourier treatment of the equations can be expressed as 
\begin{equation}
    \mathbf{k_{\perp}} = k_x\nabla x + k_y\nabla y,
\end{equation}
where $\mathbf{k_{\perp}}$ is the wavevector perpendicular to $\mathbf{B}$. Regarding the velocity coordinates, the code employs $\{v_{\|}, \mu\}$, where $v_{\|}$ is the component of the velocity parallel to the magnetic field line and $\mu = m_j v_{\perp}^2/2B$ is the magnetic moment, with $v_{\perp}$ the component of the velocity perpendicular to the magnetic field, $B=|\mathbf{B}|$ the magnetic field strength and $m_j$ the mass of the species $j$.\\

On the other hand, the direction of measurement of the DR system is perpendicular both to the radial direction and to the magnetic field {\cite{Carralero2021}}. Thus, defining the unitary vector pointing along the direction of measurement as $\mathbf{e}^{\mathrm{DR}}={\mathbf{B}\times\nabla x}/{|\mathbf{B}\times \nabla x|}$, the perpendicular wavenumber of measurement is expressed as
\begin{equation}\label{eq:k_DR}
    k_{\perp}^{\mathrm{DR}} = \mathbf{k}_{\perp}\cdot \mathbf{e}^{\mathrm{DR}} =  \mathbf{k_{\perp}}\cdot\frac{\mathbf{B}\times\nabla x}{|\mathbf{B}\times \nabla x|}.
\end{equation}
Equation (\ref{eq:k_DR}) can be written in the coordinates used in \texttt{stella} as 
\begin{equation}\label{eq:k_relation}
    k_y\rho_i = \sqrt{2m_iT_i}\frac{|\nabla x|}{eB}k_{\perp}^{\mathrm{DR}},
\end{equation}
where we have taken into account that the wavenumbers defined in \texttt{stella} are normalized with the ion thermal gyroradius $\rho_i$, given by
\begin{equation}\label{eq:rho_norm}
    \rho_i = \frac{v_{th,i}m_i}{Z_ieB_a},
\end{equation}
with $Z_i$ the ion charge number, $e$ the proton charge, $B_a$ a reference magnetic field, whose explicit expression can be found in \cite{Barnes2019, GonzlezJerez2022} and $v_{th,i}=\sqrt{2T_i/m_i}$ the ion thermal speed, where $T_i$ is the ion temperature. Expression (\ref{eq:k_relation}) implies that, given a certain set of plasma profiles and a certain wavelength of the microwave beam launched by the DR system, the corresponding value of the normalized binormal component of the wavevector defined by the code changes with the spatial location. This is due to the modification from point to point of the geometric quantities $|\nabla x|$ and $B$, as well as of $T_i(r)$.


\section{Plasma scenarios and simulation settings}\label{sec:parameters}
\begin{figure}[H]
    \centering
    \begin{subfigure}[b]{0.32\linewidth}    
        \centering
        \includegraphics[width=\linewidth]{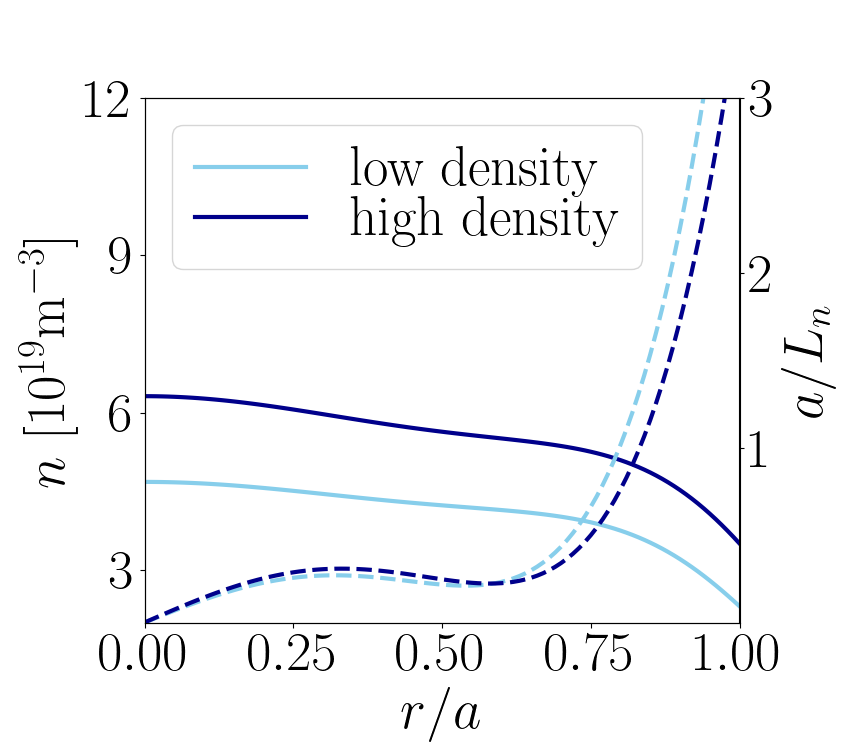}
    \end{subfigure}
    \begin{subfigure}[b]{0.32\linewidth}
        \centering
        \includegraphics[width=\linewidth]{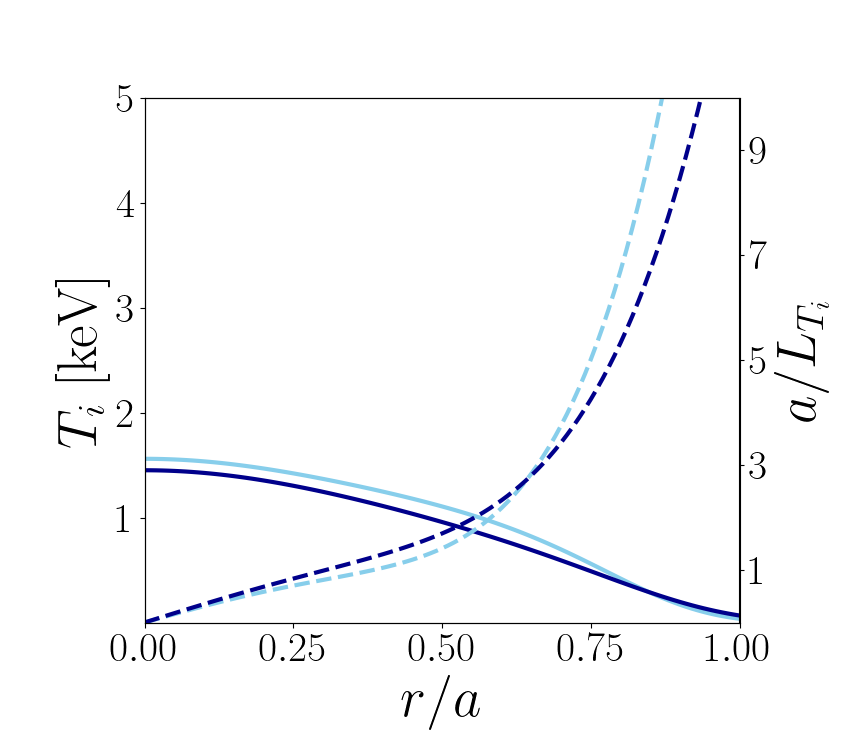}
    \end{subfigure}
    \begin{subfigure}[b]{0.32\linewidth}
        \centering
        \includegraphics[width=\linewidth]{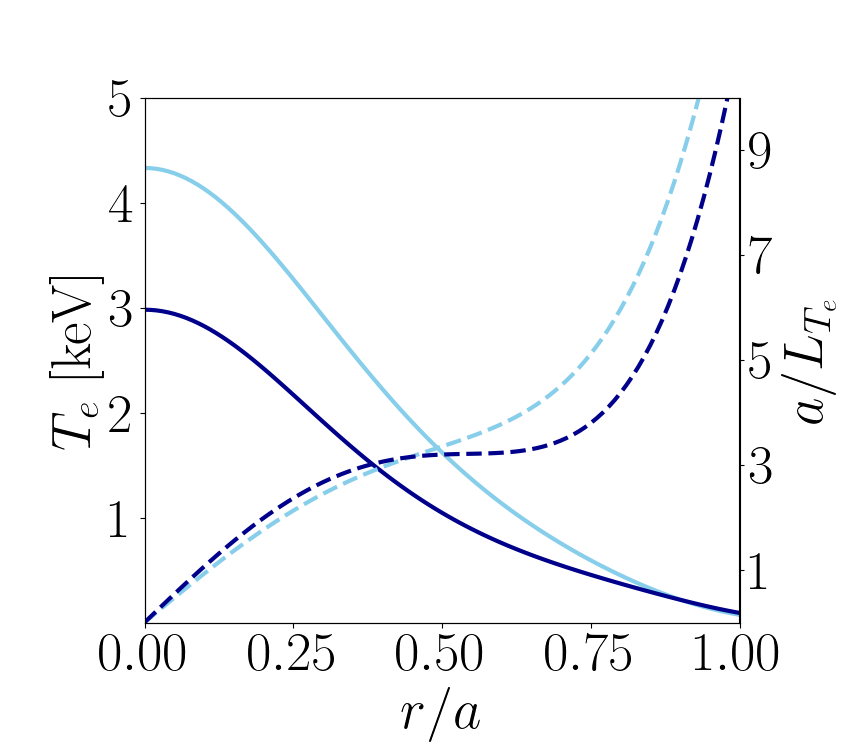}
    \end{subfigure}
    \caption{Radial profiles for the low (light blue) and high (dark blue) density scenarios. From left to right: density (solid line) and its normalized gradient (dashed line), ion temperature (solid line) and its normalized gradient (dashed line), electron temperature (solid line) and its normalized gradient (dashed line).}
    \label{fig:ECRH_profs}
\end{figure}

    In this section we describe the input for the simulations, including details about the flux tube choice, resolution, magnetic equilibrium and plasma profiles considered throughout the present work. \blue{The selected plasma parameters correspond to two gas puff-fueled ECRH scenarios, without NBI injection for heating, from the OP1.2b campaign of W7-X. They belong to the discharges \#180920.13 and \#180920.17, both using the same heating power of $P_{ECRH}\simeq 4.7$ MW without electron cyclotron current drive (ECCD)} and with the former featuring a lower density than the latter. Hence, in this paper, they will be referred to as `low density scenario' (\#180920.13) and `high density scenario' (\#180920.17). Analyses of the amplitude of the fluctuations measured with the DR system have been presented for these two discharges in \cite{Carralero2021}, which allows a straightforward comparison of the calculations obtained with \texttt{stella} and those already published. The profiles of density $(n)$, ion temperature $(T_i)$ and electron temperature $(T_e)$ of these two scenarios, obtained as fits to the experimental data, are represented in figure \ref{fig:ECRH_profs}.

\begin{figure}[H]
    \centering
    \begin{subfigure}[b]{0.35\linewidth}    
        \centering
        \includegraphics[width=\linewidth]{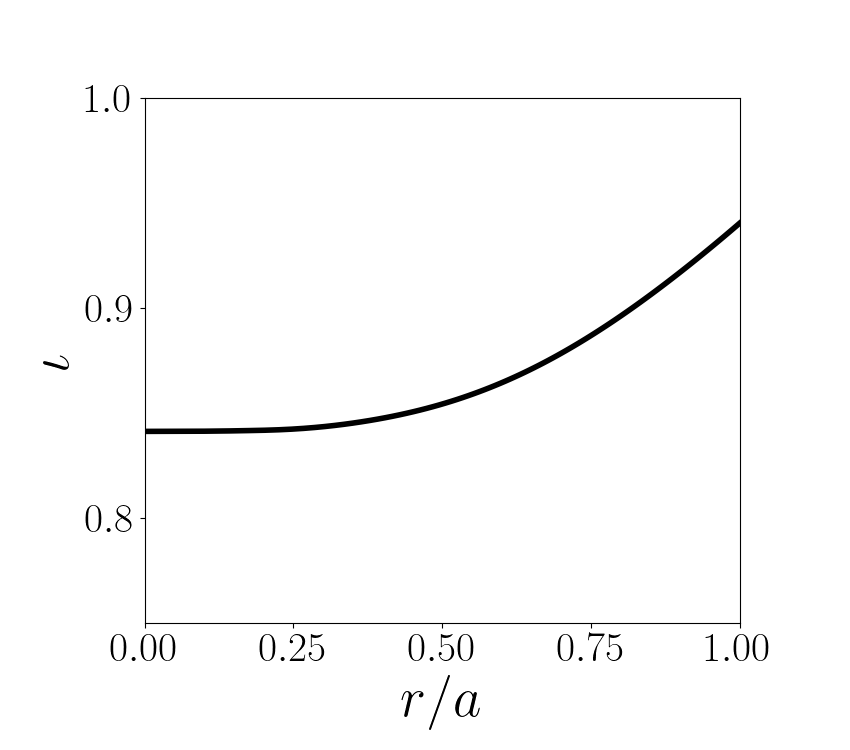}
    \end{subfigure}
    \begin{subfigure}[b]{0.35
    \linewidth}
        \centering
        \includegraphics[width=\linewidth]{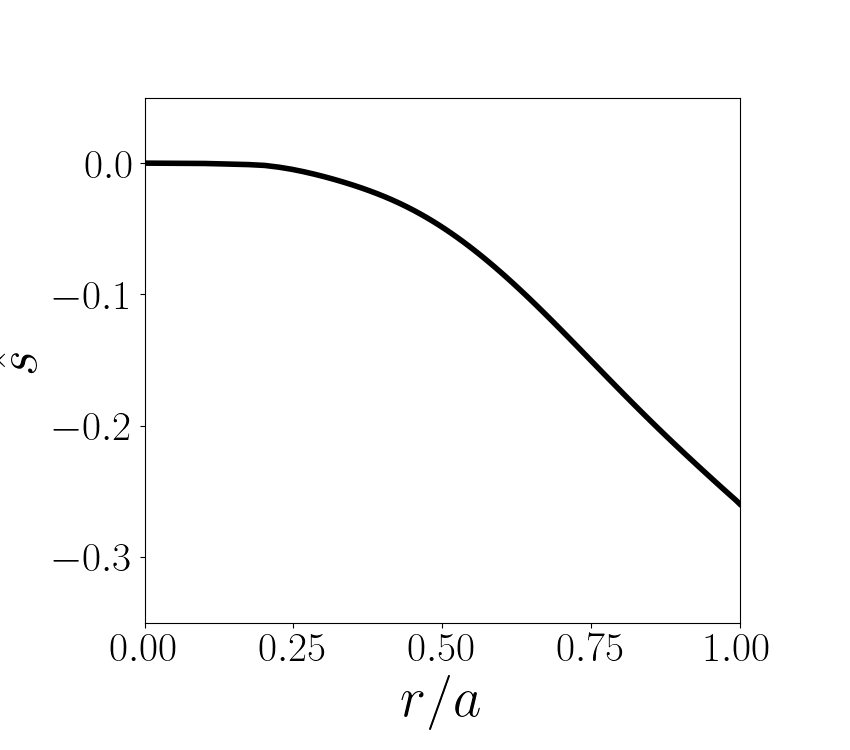}
    \end{subfigure}
    \caption{\blue{Radial profiles of $\iota$ (left) and $\hat{s}$ (right) of the standard (or EIM) configuration considered throughout this paper.}}
    \label{fig:iota}
\end{figure}

\noindent
In this figure, the normalized gradients of these quantities, defined as

\begin{align}
   a/L_{T_{i(e)}} := -a\left(\frac{\dd \ln T_{i(e)}}{\dd r}\right), &&    a/L_{n} := -a\left(\frac{\dd \ln n}{\dd r}\right),
\end{align}
are also depicted. In order to cover the radial range of measurement of the DR system reported in \cite{Carralero2021}, five radial positions, labeled by $r_0/a=\{0.5,0.6,0.7,0.8,0.9\}$, have been considered. The values of the plasma profiles and their normalized gradients for the selected radial locations are listed in table \ref{tab:parameters}. Finally, the magnetic geometry used as input for the simulations is provided by the ideal MHD equilibrium code \texttt{VMEC} \cite{Hirshman_1986}. This configuration is an example of a standard (\blue{also known as EIM}) W7-X equilibrium \blue{in which all planar coil currents are set to zero and the non-planar coil current are set to the same value} (see \cite{Geiger2015} for a detailed description of the different W7-X configurations) with \blue{effective minor radius $a=0.514$ m, major radius $R_0=5.513$ m}, magnetic field on-axis $B_{ax}=2.52 \; \mathrm{T}$, and, for normalization purposes, see equation (\ref{eq:rho_norm}), $B_a = 2.28 \; \mathrm{T}$. \blue{The global shear, defined as $\hat{s}= - \frac{r}{\iota}\frac{\dd \iota}{\dd r}$, and $\iota$ profiles corresponding to this configuration are represented in figure \ref{fig:iota}}.

\begin{figure}[H]
    \centering
    \begin{subfigure}[b]{0.5\linewidth}        
        \centering
        \includegraphics[width=\linewidth]{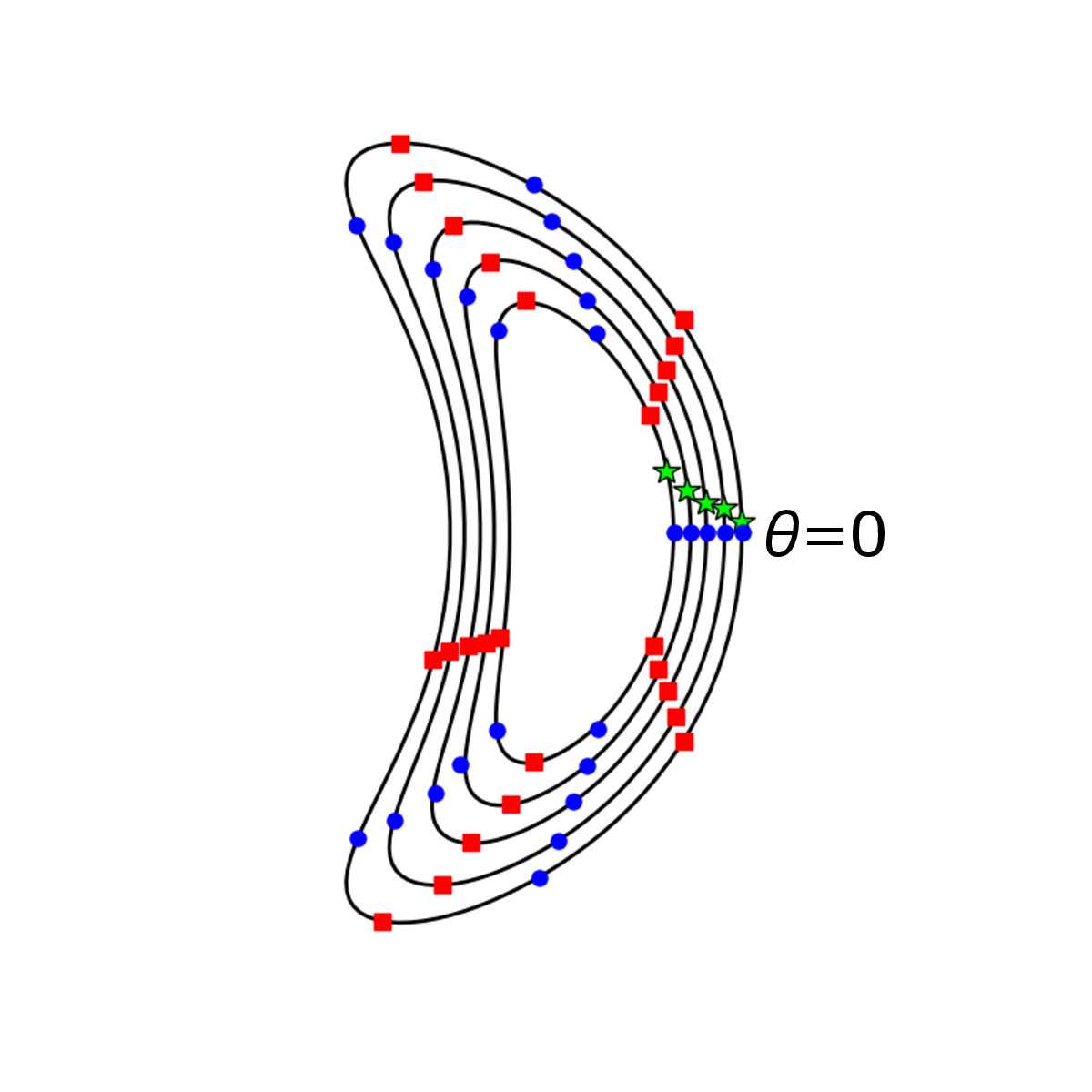}
    \end{subfigure}
    \caption{Points belonging to the bean (blue dots) and triangular (red squares) flux tubes of W7-X that lay on the plane $\zeta = 0$. The measurement positions for the high density scenario, obtained with \texttt{TRAVIS} for the radial locations considered in the gyrokinetic simulations, are represented as stars.}
    \label{fig:DR_view}
\end{figure}

\begin{figure}[H]
    \centering
    \begin{subfigure}[b]{0.48\linewidth}        
        \centering
         \includegraphics[width=\linewidth]{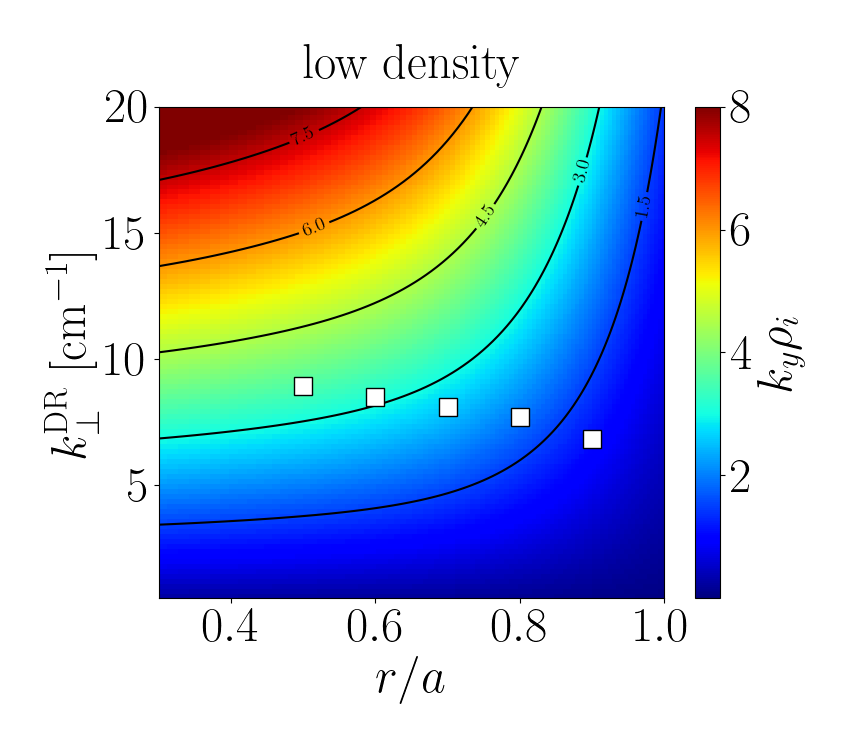}
    \end{subfigure}
\begin{subfigure}[b]{0.48\linewidth}        
        \centering
        \includegraphics[width=\linewidth]{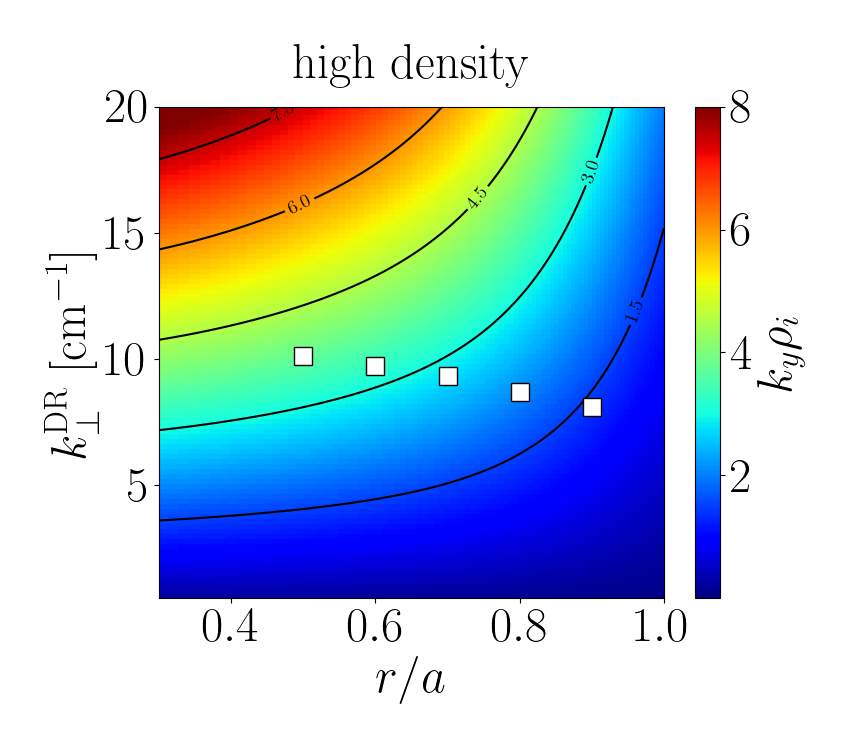}
    \end{subfigure}
    \caption{Relation between $k_y\rho_i$ and $k_{\perp}^{\mathrm{DR}}$ as a function of the radial position for a wide range of $k_{\perp}^{\mathrm{DR}}$. These plots are the result of evaluating expression (\ref{eq:k_relation}) at $(\theta, \zeta) = (0,0)$ considering the $T_i$ profiles represented in figure \ref{fig:ECRH_profs} for the low (left) and high (right) density scenarios. The white squares represent the perpendicular wavenumbers measured by the DR system at each radial location.}
    \label{fig:k_relation}
\end{figure}

{During the first W7-X operational phase (OP1), a DR system measured at locations nearby the bean-shaped toroidal plane $(\zeta = 0)$ of the plasma.} The system launches a microwave beam whose cut-off determines the point of measurement of the perpendicular flow of density fluctuations. Those positions can be obtained applying ray-tracing techniques. In figure \ref{fig:DR_view}, the measurement positions for the high density scenario\footnote{The measurement positions for the low density scenario are very close to those represented in figure \ref{fig:DR_view}.} at the radial locations considered in the gyrokinetic simulations, obtained using the ray-tracing code \texttt{TRAVIS} \cite{Marushchenko2014}, are represented with stars. The intrinsic restrictions of the flux tube formalism and the boundary conditions used in this study to access those measurement locations should also be noted. The parallel boundary conditions implemented for this work \cite{Martin2018} restrict the selected flux tubes to stellarator symmetric ones, i.e. those fulfilling $B(\theta, \zeta)=B(-\theta, -\zeta)$. In W7-X, the suitable flux tubes are then the ‘bean’ flux tube, which is centered with respect to the position $(\theta,\zeta) = (0,0)$ and the ‘triangular’ flux tube, which is centered at $(\theta,\zeta) = (0,\pi/5)$ \blue{\cite{GonzlezJerez2022}}. The intersections of these two flux tubes, assuming they extend one turn in the poloidal direction, with the bean-shaped toroidal plane of W7-X can also be found in figure \ref{fig:DR_view}. The bean flux tube position $(\theta,\zeta) = (0,0)$ provides the spatial location nearest to the positions of measurements of the DR system. Therefore, the information to compare against the measurements of the DR system will be extracted from \blue{that flux tube at $(\theta,\zeta) = (0,0)$}. In addition, concerning the perpendicular wavenumber of measurement, the {relationship} between $k_{\perp}^{\mathrm{DR}}$ and $k_y\rho_i$ (see expression (\ref{eq:k_relation})) is represented as a function of the radial coordinate in figure \ref{fig:k_relation} for a wide range of $k_{\perp}^{\mathrm{DR}}$. The $k_{\perp}^{\mathrm{DR}}$ values selected in the experiment (included in table \ref{tab:parameters} and reported in \cite{Carralero2021}) are indicated in figure \ref{fig:k_relation} with white squares. Finally, the simulations performed in this work have considered a phase space grid of $\{92 \times 36 \times 12 \times 59 \times 45\}$ points in $\{\zeta, v_{\|}, \mu, k_x, k_y\}$ and maximum and minimum values of $k_x$ and $k_y$, $(k_x^{\mathrm{max}}\rho_i, k_y^{\mathrm{max}}\rho_i)\simeq (2.5, 4.5)$, {and $(k_x^{\mathrm{min}}\rho_i, k_y^{\mathrm{min}}\rho_i) \simeq (0.1,0.1)$, respectively}.

\begin{table}[H]
\centering
\begin{tabular}{c | c | c c c c c c c}
    \toprule
  $r_0/a$ & discharge & $n$ $[10^{19}\mathrm{m}^{-3}]$ & $a/L_{n}$ & $T_i$ [keV] & $a/L_{T_i}$ & $T_e$ [keV] & $a/L_{T_e}$ & $k_{\perp}^{\mathrm{DR}}$ $[\mathrm{cm^{-1}}]$ \\ \midrule \midrule
   \multirow{2}{*}{0.5} &
   low  & 4.23 & 0.22 & 1.11 & 1.41 & 1.62 & 3.35 & 8.9 \\
   &high & 5.63 & 0.25 & 0.96 & 1.70 & 1.04 & 3.20 & 10.1\\ \midrule
   \multirow{2}{*}{0.6} & 
   low & 4.14 & 0.23 & 0.93 & 2.18 & 1.13 & 3.78 & 8.5 \\
   &high & 5.51 & 0.22 & 0.78 & 2.33 & 0.75 & 3.23 & 9.7\\ \midrule
   \multirow{2}{*}{0.7} & 
   low & 4.01 & 0.43 & 0.70 & 3.75 & 0.75 & 4.51 & 8.1 \\
   &high & 5.36 & 0.34 & 0.59 & 3.43 & 0.54 & 3.47 & 9.3\\ \midrule
   \multirow{2}{*}{0.8} & 
   low & 3.75 & 1.02 & 0.42 & 6.70 & 0.45 & 5.97 & 7.7 \\
   &high & 5.09 & 0.76 & 0.39 & 5.32 & 0.37 & 4.37 & 8.7\\ \midrule
   \multirow{2}{*}{0.9} & 
   low & 3.21 & 2.25 & 0.17 & 11.81 & 0.22 & 8.73 & 6.8 \\
   &high & 4.52 & 1.73 & 0.20 & 8.48 & 0.22 & 6.63 & 8.1\\ \bottomrule
\end{tabular}
   \caption{Local parameters used in the simulations at each radial position for the low and high density scenarios. From left to right: density, normalized density gradient, ion temperature, normalized ion temperature gradient, electron temperature, normalized electron temperature gradient and perpendicular wavenumber measured by the DR system.}
   \label{tab:parameters}
\end{table}

\section{Radial dependence of density fluctuations: comparison between \texttt{stella} results and DR measurements}\label{sec:comparisons}
\begin{figure}[H]
    \centering
        \includegraphics[width=0.5\linewidth]{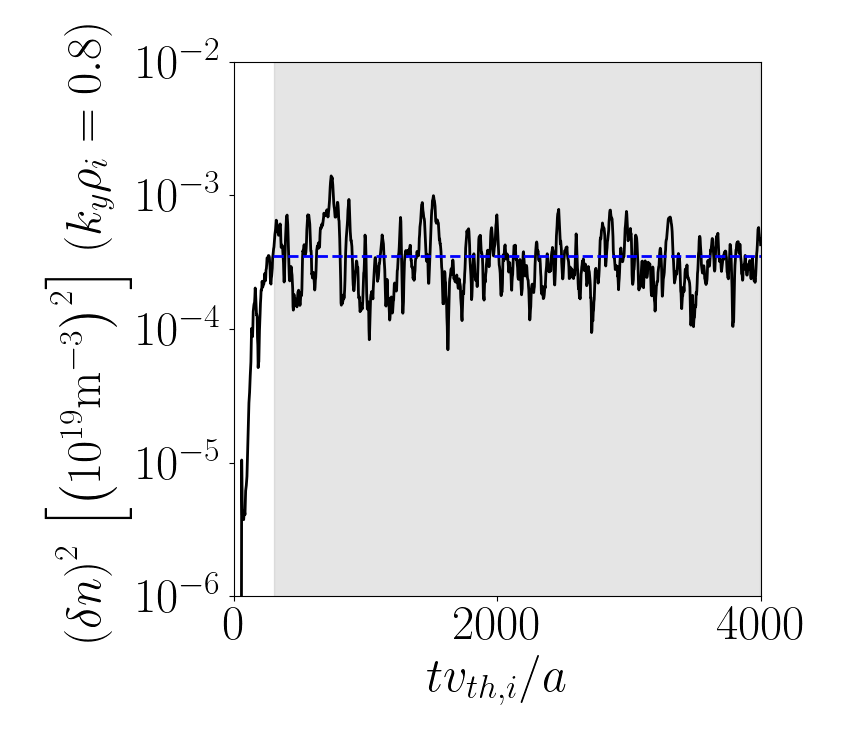}
    \caption{Time evolution of $(\delta n)^2(k_y\rho_i = 0.8)$ at $r_0/a=0.6$ in the high density scenario. The range considered for the time-average is represented by the shadowed area and the mean value of this quantity is indicated with the horizontal dashed line.}
    \label{fig:dens_t}
\end{figure}

    The DR system measures the backscattered power $(S)$ of a microwave beam launched into the plasma, which, for low turbulence levels, is proportional to the squared amplitude of the density fluctuations $\left(S \propto |\delta n|^2\right)$ \cite{Gusakov2004}. On the other hand, the code \texttt{stella} provides, at each instant and each position along the flux tube, the decomposition in Fourier space of the density fluctuations $(\delta n)$ on the plane perpendicular to the magnetic field. The density fluctuations at $r_0$ and $\alpha_0$ can be expressed as
\begin{equation}
    \delta n\left(x, y, \zeta, t\right) = \sum_{k_x,k_y}\widehat{\delta n}_{k_x,k_y}(\zeta, t)\exp[\mathrm{i}(k_x x + k_y y)].
\end{equation}
As explained in section \ref{sec:parameters}, the simulated positions nearest to the measurement locations are those of the bean flux tube with $\zeta=0$. Once $|\widehat{\delta n}_{k_x,k_y}|^2(\zeta=0, t)$ is obtained from the simulations, the quantity to compare against the DR system measurements is post-processed as follows. First, for each pair of wavenumbers $(k_x,k_y)$, the spectrum in $k_y$ is obtained by summing over $k_x$ as
\begin{equation}\label{eq:delta_n_DR}
    \left(\delta n\right)^2(k_y, t) = \sum_{k_x}\abs{\widehat{\delta n}_{k_x,k_y}}^2 (\zeta=0,t).
\end{equation}
Second, we time-average each $(\delta n)^2(k_y, t)$ over the saturated nonlinear phase, obtaining the time-averaged squared density fluctuations, $\langle (\delta n)^2\rangle_t(k_y)$. For illustrative purposes, figure \ref{fig:dens_t} shows the time trace of $(\delta n)^2$ for the mode $k_y\rho_i = 0.8$, which is the one with the largest amplitude at $r_0/a=0.6$ in the high density scenario.

\begin{figure}[H]
    \centering
    \begin{subfigure}[b]{0.48\linewidth}        
        \centering
        \includegraphics[width=\linewidth]{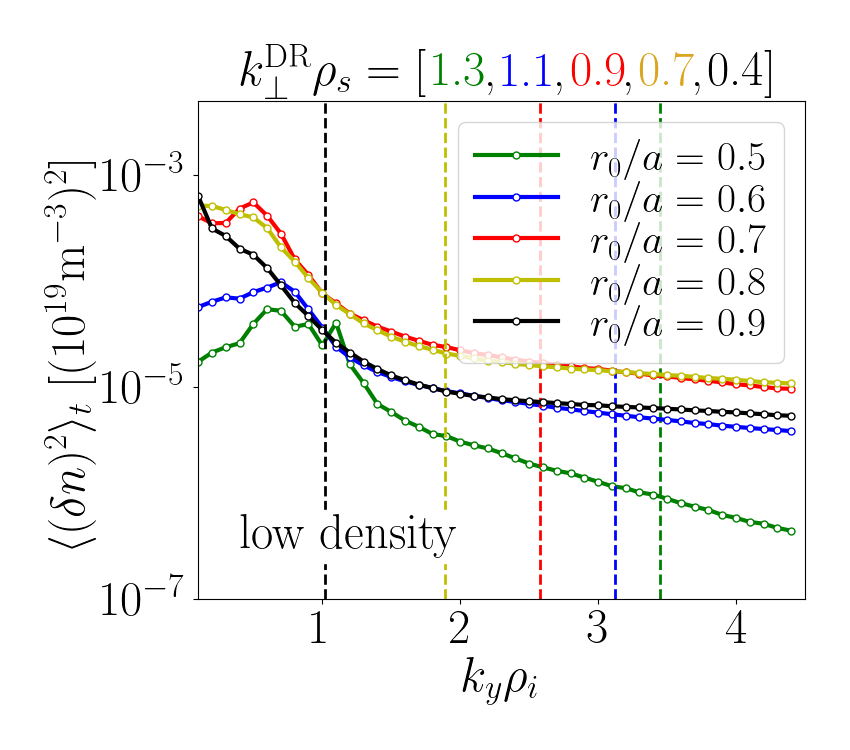}
    \end{subfigure}
\begin{subfigure}[b]{0.48\linewidth}        
        \centering
        \includegraphics[width=\linewidth]{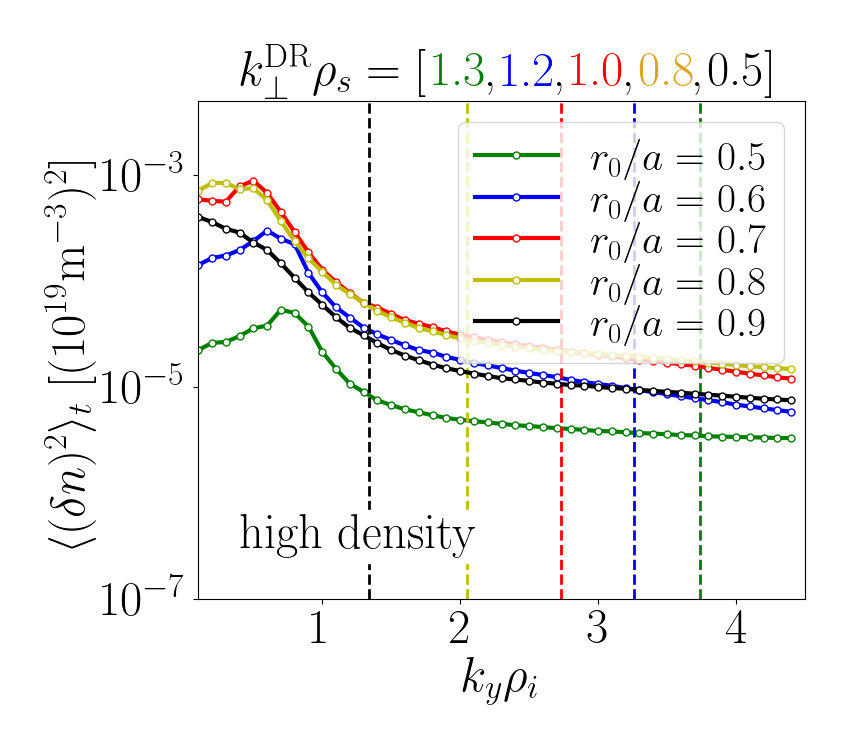}
    \end{subfigure}
    \caption{$k_y$ spectrum of the time-averaged squared density fluctuations, $\langle(\delta n)^2\rangle_t$, obtained with \texttt{stella} for the low (left) and high (right) density scenarios. {The dashed vertical lines indicate the wavenumber measured by the DR system, $k^{\mathrm{DR}}_{\perp}$, at each radial position. The values of $k^{\mathrm{DR}}_{\perp}$ are normalized following the convention employed in [18], where $\rho_s:=\sqrt{T_im_i}/eB$.}}
    \label{fig:dens_ky}
\end{figure}

In figure \ref{fig:dens_ky}, the results obtained for $\langle(\delta n)^2\rangle_t$ for the five scanned radial positions and for the two scenarios are represented as a function of $k_y\rho_i$. The values of $k_y$ that correspond to the wavenumbers of measurement of the DR system, that we denote by $k_y^{\mathrm{DR}}=k_y(k_{\perp}^{\mathrm{DR}})$, are also indicated with vertical dashed lines in these plots. It can be observed that the DR system, depending on the radial location, measures very disparate values of $k_y\rho_i$. Whereas at $r_0/a=0.9$ the system accesses nearly scales with $k_y\rho_i\sim 1$, as $r_0$ decreases it explores different regions of the spectrum, reaching up to $k_y\rho_i\sim 3.5$. 

\begin{figure}
    \centering
    \begin{subfigure}[b]{0.48\linewidth}        
        \centering
        \includegraphics[width=\linewidth]{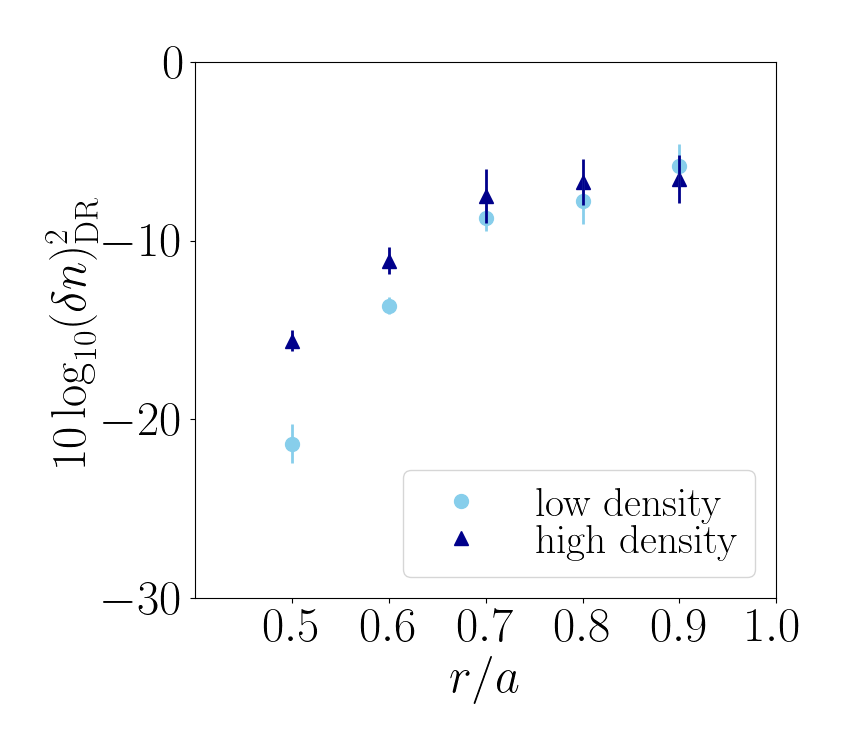}
    \end{subfigure}
\begin{subfigure}[b]{0.48\linewidth}        
        \centering
        \includegraphics[width=\linewidth]{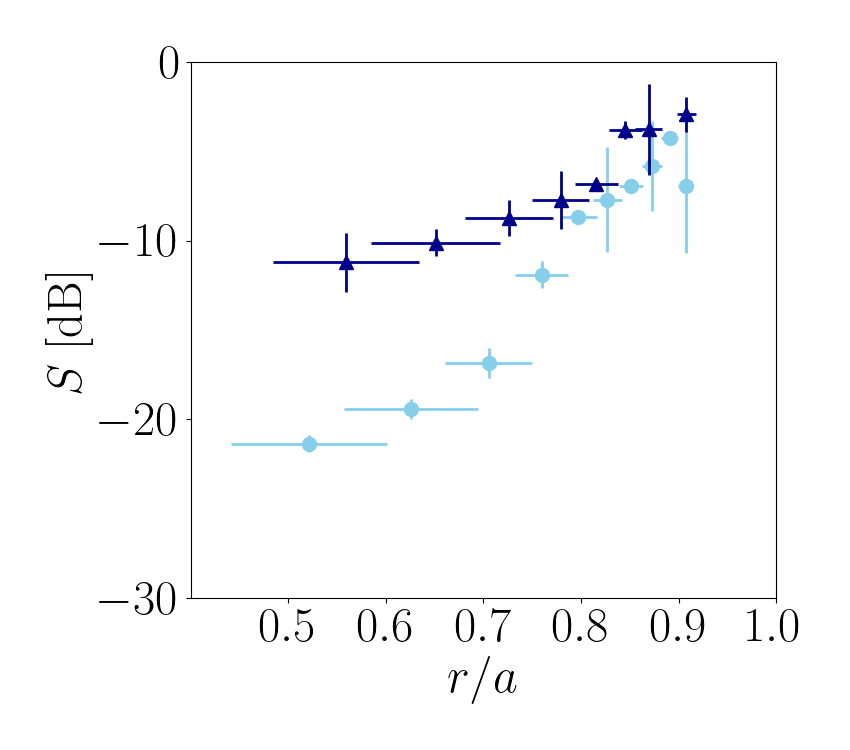}
    \end{subfigure}
    \caption{Squared amplitude of the density fluctuations as a function of the radial coordinate for the low (light blue circles) and high (dark blue triangles) density scenarios. Left: Numerical results of $(\delta n)^2_{\mathrm{DR}}$ obtained with \texttt{stella}. The error bars represent the standard deviation from the mean value evaluated over the saturated nonlinear phase. Right: backscattered power measured by the DR system.}
    \label{fig:dens_r_DR}
\end{figure}

\begin{figure}
    \centering
    \begin{subfigure}[b]{0.48\linewidth}        
        \centering
        \includegraphics[width=\linewidth]{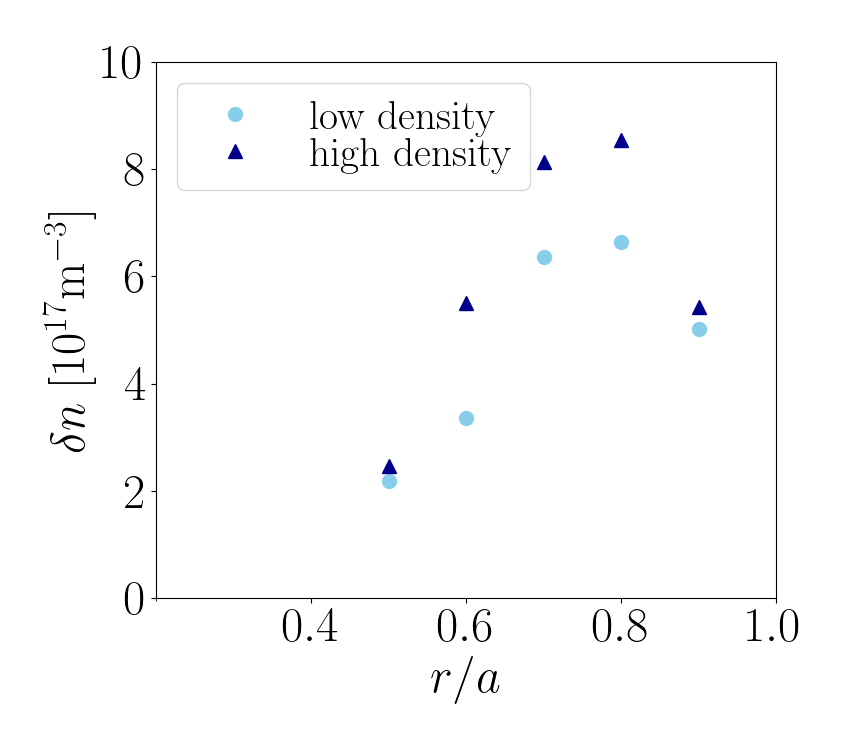}
    \end{subfigure}
    \caption{Amplitude of the density fluctuations integrated over $k_x$ and $k_y$ computed with \texttt{stella} as a function of the radial coordinate for the low (light blue circles) and high (dark blue triangles) density scenarios.}
    \label{fig:dens_total_r}
\end{figure}

Finally, considering the squared density fluctuations for each radial location at the specific value of $k_y$ accessed by the diagnostic,
\begin{equation}
    (\delta n)^2_{\mathrm{DR}} = \langle(\delta n)^2\rangle_t (k_y=k_y^{\mathrm{DR}}),
\end{equation}
the comparison against the backscattered power, $S$, can be carried out. In figure \ref{fig:dens_r_DR} (left), the numerical results obtained with \texttt{stella}, $(\delta n)^2_{\mathrm{DR}}$, are represented as a function of the radial coordinate, while in figure \ref{fig:dens_r_DR} (right), the measurements of $S$ obtained with the DR system (reported in \cite{Carralero2021}) can be found. It is important to note that the DR measurements represented in figure \ref{fig:dens_r_DR} (right) are expressed in dB, but the reference normalization value is unknown. Hence, $S$ [dB] $\propto 10\log_{10}(\delta n)^2_{\mathrm{DR}} + C$, with $C$ an arbitrary constant. For this comparison, $C$ has been chosen to make $ S\left(r_0/a=0.5\right) = 10\log_{10}(\delta n)^2_{\mathrm{DR}}\left(r_0/a=0.5\right) + C$ or, equivalently, to set the lowest value of the numerical and experimental results at the same level.
It can be observed that the numerical results shown in figure \ref{fig:dens_r_DR} (left) and the measurements in figure \ref{fig:dens_r_DR} (right) exhibit a monotonic increase of the squared density fluctuations with $r/a$. In addition, the difference between their minimum and maximum values is approximately 15 dB in both cases. Finally, the numerical and experimental results corresponding to the low density scenario {show lower} turbulent fluctuations than those of the high density scenario. These features, in particular the monotonic increase of the density fluctuations towards the edge, result, {partially},  from the fact that the DR system measures, as the radial position changes, at different locations of the $k_y$ spectrum, as we anticipated in figure \ref{fig:dens_ky} and related discussion. Indeed, the density fluctuations integrated both in $k_x$ and $k_y$, i.e $\delta n = \sqrt{\sum_{k_y}\langle(\delta n)^2\rangle_t}$, loses the monotonic increasing behaviour with $r/a$, as figure \ref{fig:dens_total_r} illustrates, for the two discharges analyzed. In that figure, one can observe that the integrated density fluctuations have a maximum at nearly $r_0/a=0.8$ and ranges between $2$ and $8\times 10^{19}$ m$^{-3}$. This picture aligns remarkably well with what Phase Contrast Imaging (PCI) techniques and \texttt{GENE} \cite{Jenko2000} simulations have reported for similar plasmas in W7-X \cite{Bhner2021}.
On the other hand, returning to the comparison between \texttt{stella} calculations and DR measurements of figure \ref{fig:dens_r_DR}, the numerical results do not exhibit the significant reduction that the measurements do for the low density scenario in the range $r/a \sim [0.6, 0.7]$. A possible source of discrepancy could be that the bean flux tube at $\zeta=0$ does not correspond to the exact spatial measurement position of the DR system (see figure \ref{fig:DR_view}). Nevertheless, as it is explained in \ref{ap:dens_poloidal}, when the poloidal deviation from the bean flux tube at $\zeta=0$ is considered, the numerical results do not change significantly. 
\blue{With respect to the model used, possible extensions, such that accounting for the full flux surface geometry \cite{CarraleroISHW2022} and, in that case, including the radial electric field as well, might be worth addressing in order to attempt a better agreement.} \red{With regard to collisions, which have been neglected, the normalized ion collision frequency between $r/a=0.5-0.8$ (considering ions and electrons with one thermal speed) is $a\nu_i/v_{th,i}\approx 1.3\times 10^{-3}-1\times 10^{-2}$. This value is significantly smaller than the growth rates of most unstable electrostatic instabilities, typically of a few tenths in units of $a/v_{th,i}$. Exceptionally, at the position $r/a=0.9$, $a\nu_i/v_{th,i}\approx 3.5\times 10^{-2}$ in both scenarios. Electron collisions are higher though ($a\nu_e/v_{th,i}\approx 0.1-3$, with the lowest values at $r/a=0.5$), nevertheless  electrons do not seem to play a prominent role except at the innermost studied position (see section 6). The impact of these factors will be addressed in forthcoming works.}


\section{Numerical characterization of fluctuations in the frequency domain}\label{sec:predictions}
        This section extends the analysis of the fluctuations, performed for the two W7-X discharges described in section \ref{sec:parameters}, to the frequency domain. The aim is to complete the characterization of the fluctuations with features that can be inferred from the time evolution of $S$ and $u_{\perp}$ measured by the DR system. In particular, these features correspond to the frequency spectrum of the squared amplitude of density fluctuations, presented in section \ref{subsec:spectra_density}, and the frequency spectrum of the zonal component of the $E\times B$ flow, discussed in section \ref{subsec:spectra_ExB}. Future analyses, that automate the measurement of highly time-resolved traces of $S$, will enable a straightforward comparison against the fluctuation frequency spectra provided in section \ref{subsec:spectra_density}. On the other hand, the installation in a different toroidal sector of a second DR system ---whose data acquisition and analysis will extend during the second W7-X operation phase (OP2)--- will allow to identify oscillations in $u_{\perp}$ of zonal origin to be verified against the spectra presented in section \ref{subsec:spectra_ExB}. \red{For these future comparisons it will be necessary to subtract the Doppler shift from the experimental measurements of frequency spectra.}
    \subsection{Frequency spectra of density fluctuations}\label{subsec:spectra_density}
\begin{figure}[H]
    \centering
    \includegraphics[width=\linewidth]{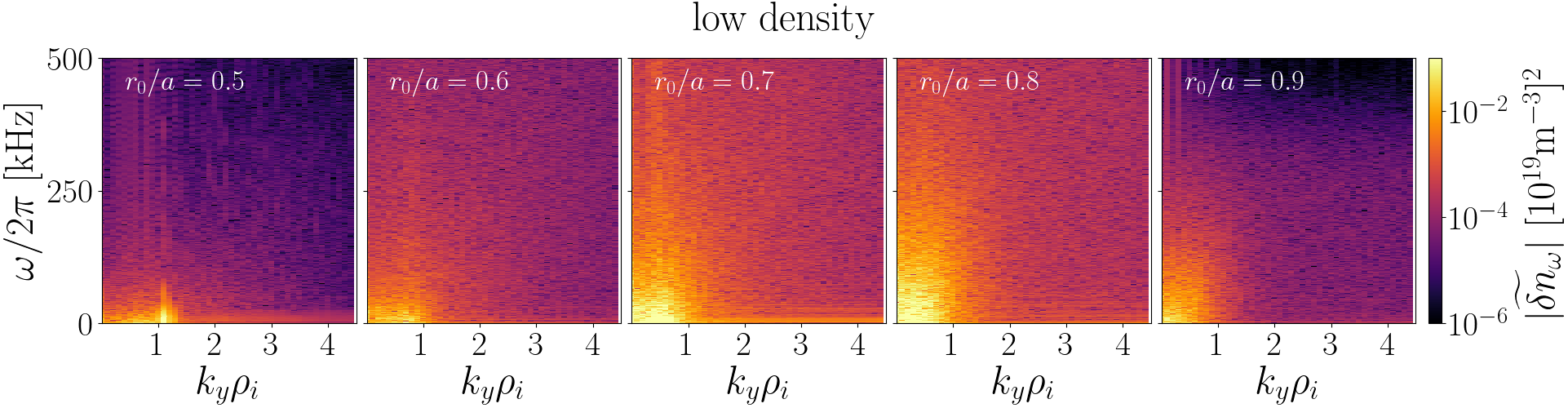}
    \includegraphics[width=\linewidth]{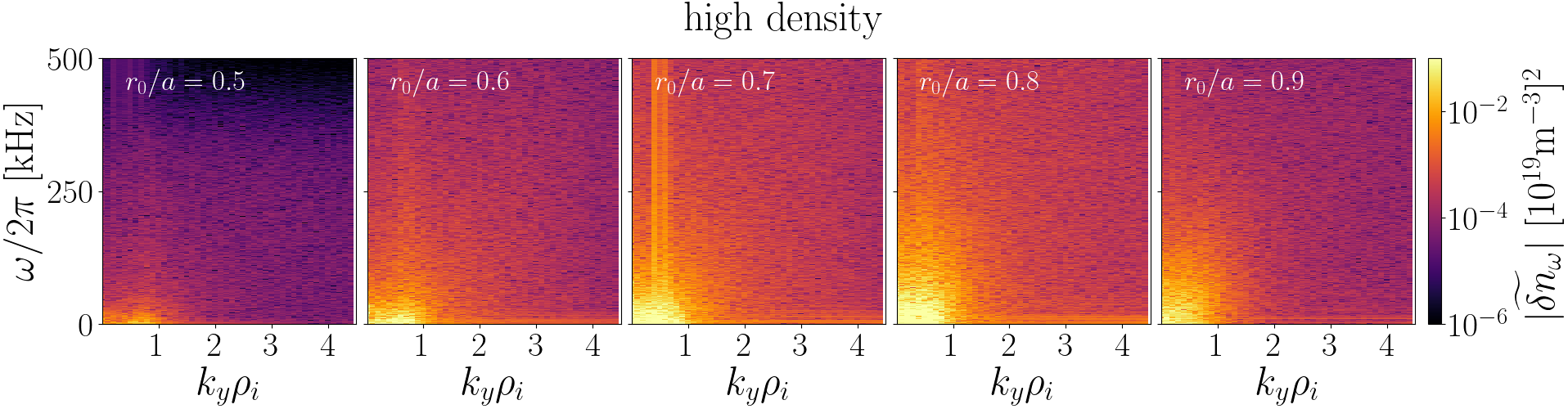}
    \caption{Frequency spectra of $(\delta n)^2$ for the low (top) and high (bottom) density scenarios computed with \texttt{stella} as function of $k_y$. }
    \label{fig:dens_freq_ampl}
\end{figure}

\begin{figure}[H]
    \centering
    \includegraphics[width=\linewidth]{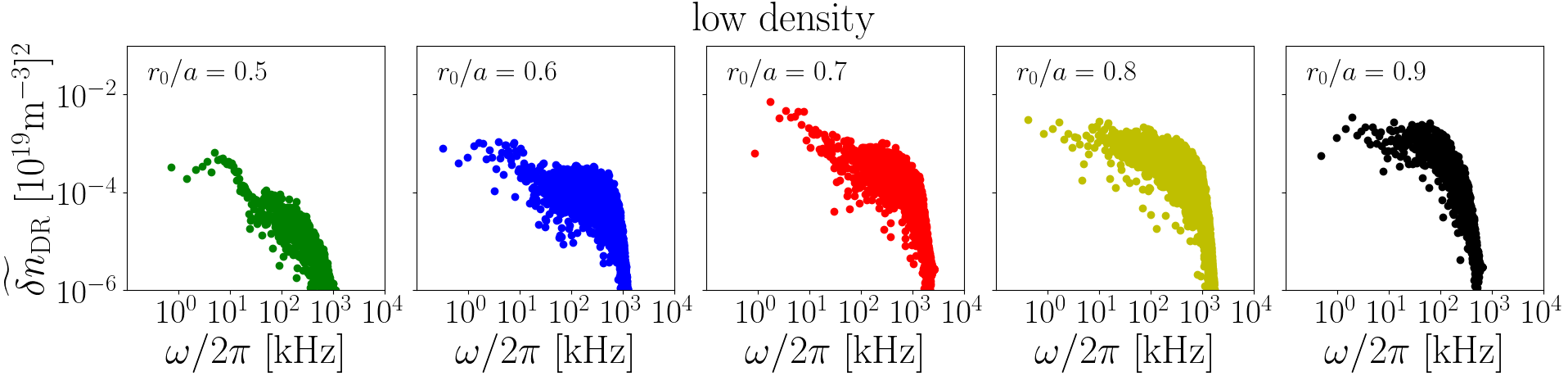}
    \includegraphics[width=\linewidth]{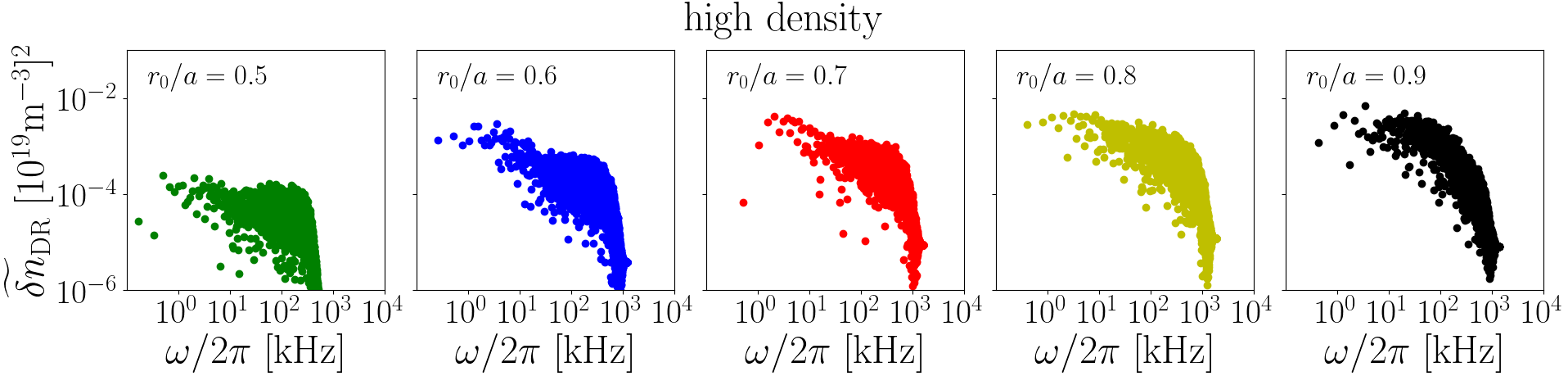}
    \caption{{Density spectra of} $\widetilde{\delta n}_{\mathrm{DR}}$ computed with \texttt{stella} for the low (top) and high (bottom) density scenarios {for the case $k_y = k_y^{\mathrm{DR}}$}.}
    \label{fig:dens_freq_ampl_DR}
\end{figure}

        In section \ref{sec:comparisons}, we have addressed the comparison between the radial dependence of the squared amplitude of the density fluctuations computed with \texttt{stella} and those measured by the DR system. In that case, a time average of $(\delta n)^2(k_y, t)$ was performed. If, instead, the time dependent squared density fluctuations, $(\delta n)^2(k_y,t)$, are considered and a Fourier transform in $t$ is taken, we can write
\begin{equation}
        (\delta n)^2(k_y,t) = \sum_{\omega} \widetilde{\delta n}_{\omega}(k_y)\exp(-\mathrm{i}\omega t),
\end{equation}
leading to the frequency spectrum of the squared density fluctuations. \red{Note that the obtained frequency spectrum differs from the power spectral density of $\delta n$. In this section, we assume that once the time evolution of the backscattered power $S(t)$ of the beam launched by the system is known ---since $S(t)$ is proportional to the $(\delta n)^{2}$--- the comparison of the frequency spectra of the experimental signal $S$ and the numerically obtained $(\delta n)^2$ is straightforward.} In figure \ref{fig:dens_freq_ampl}, the amplitude of $\widetilde{\delta n}_{\omega}$ is represented as a function of the frequency $\omega$ and $k_y$. We define the amplitude of the Fourier harmonics $\widetilde{\delta n}_{\omega}$ for $k_y=k_y^{\mathrm{DR}}$ as
\begin{equation}\label{eq:Fourier_dens_DR}
    \widetilde{\delta n}_{\mathrm{DR}}(\omega) = \abs{\widetilde{\delta n}_{\omega}}(k_y = k_y^{\mathrm{DR}}).
\end{equation}
\noindent
Figure \ref{fig:dens_freq_ampl_DR} shows $\widetilde{\delta n}_{\mathrm{DR}}(\omega)$ for the two scenarios under consideration. In general, for all radial positions, the spectrum is broad and extends a few hundreds of kHz. Above frequencies of that order of magnitude, an abrupt drop of the amplitude is observed. Below $\omega/2\pi\sim 100$ kHz, the spectra are rather flat. However, for some radial positions ---$r_0/a=\{0.5,0.6,0.7\}$ in the low density scenario and $r_0/a=\{0.6,0.7\}$ in the high density one---  peak values of the amplitude are observed for frequencies such that $\omega/2\pi \lesssim 10$ kHz. After a similar Fourier analysis of $S(t)$, the experimental frequency spectra could be compared against those represented in figure \ref{fig:dens_freq_ampl_DR}.

    \subsection{Frequency spectra of zonal flow fluctuations}\label{subsec:spectra_ExB}
        Zonal perturbations of the electrostatic potential $\varphi$ are constant on flux surfaces and have radial structure. In the flux tube scheme, they correspond to the components of the potential with $k_y=0$ and finite $k_x$. With a second Doppler reflectometer looking on the same flux surface as the first one but at a different toroidal location, it would be possible to identify zonal oscillations. Specifically, common frequencies found in the time evolution of $u_{\perp}(t)$ obtained by each DR could be identified with zonal $E\times B$ flow fluctuations projected along the measurement direction of the system, $\mathbf{e}^{\mathrm{DR}}$. In this subsection we address the characterization of these frequencies.\\

The fluctuating $E\times B$ velocity can be written as
\begin{equation} \label{eq:ExB}
    \delta\mathbf{{v}}_{E} = \frac{1}{B^2}\mathbf{\mathbf{B}}\times\nabla\varphi,
\end{equation}
where the fluctuating electrostatic potential can be expressed, for each radial position $r_0$, in Fourier series as 
\begin{equation}
    \varphi(x, y, \zeta, t) = \sum_{k_x,k_y}\widehat{\varphi}_{k_x, k_y}(\zeta, t)\exp[\mathrm{i}(k_x x + k_y y)].
\end{equation} 

\noindent
\green{
We consider the modes of the potential with $k_y=0$, for which expression (\ref{eq:ExB}) becomes
\begin{equation}\label{eq:ExB_ZF}
    \delta \mathbf{v}_{E}(x,\zeta,t) = \frac{\left(\mathbf{B}\times \nabla x\right)}{B^2}\sum_{k_x}\mathrm{i}k_x\widehat{\varphi}_{k_x, k_y=0}(\zeta, t)\exp(\mathrm{i}k_x x).
\end{equation}
Considering the projection of $\delta \mathbf{v}_{E}$ along the direction of measurement of the DR diagnostic, $\mathbf{e}^{\mathrm{DR}}$, we can compute the zonal flow fluctuations expected to contribute to the total perpendicular flow measured by the DR system, $\delta u_{\perp}^{\mathrm{ZF}} = \langle\delta \mathbf{v}_{E} \cdot \mathbf{e}^{\mathrm{DR}}\rangle_{\zeta}$\footnote{Here, the field line average is defined as $\langle \cdot\rangle_{\zeta}=\int_{\zeta_{\mathrm{min}}}^{\zeta_{\mathrm{max}}} (\hat{\mathbf{b}}\cdot\nabla\zeta)^{-1}(\cdot)\dd \zeta/ \int_{\zeta_{\mathrm{min}}}^{\zeta_{\mathrm{max}}} (\hat{\mathbf{b}}\cdot\nabla\zeta)^{-1}\dd \zeta$.}. Explicitly 
\begin{equation}
    \delta u_{\perp}^{\mathrm{ZF}}(x,t) =  \mathrm{i} \sum_{k_x} k_x \left<\frac{|\nabla x|}{B}\widehat{\varphi}_{k_x, k_y=0}(\zeta, t)\right>_{\zeta} \exp(\mathrm{i}k_x x),
\end{equation}
}

\begin{figure}[H]
    \centering
    \includegraphics[width=\linewidth]{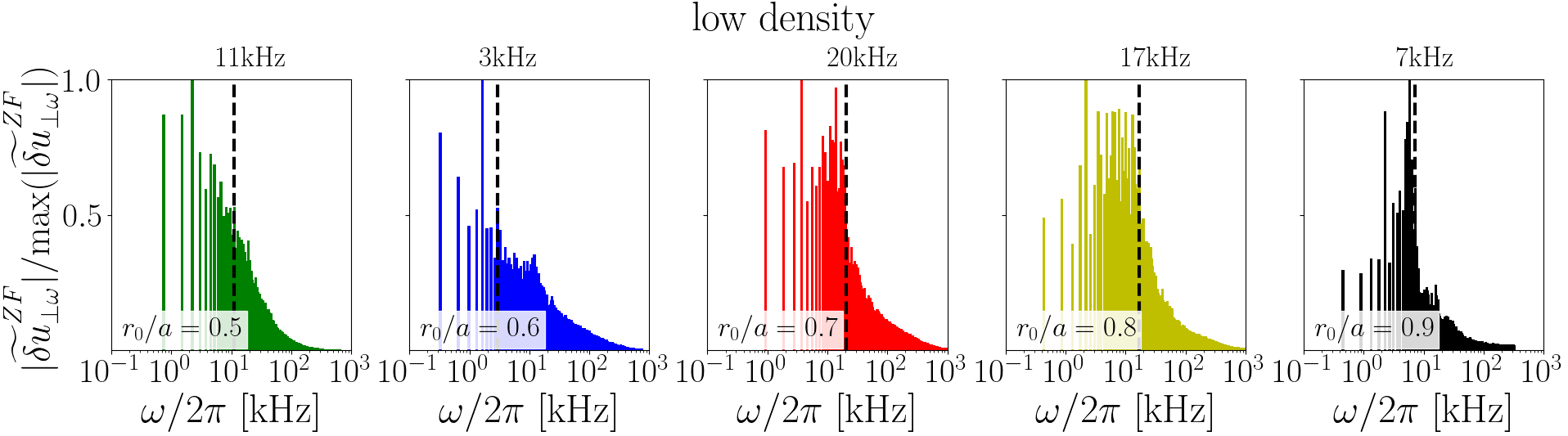}
    \includegraphics[width=\linewidth]{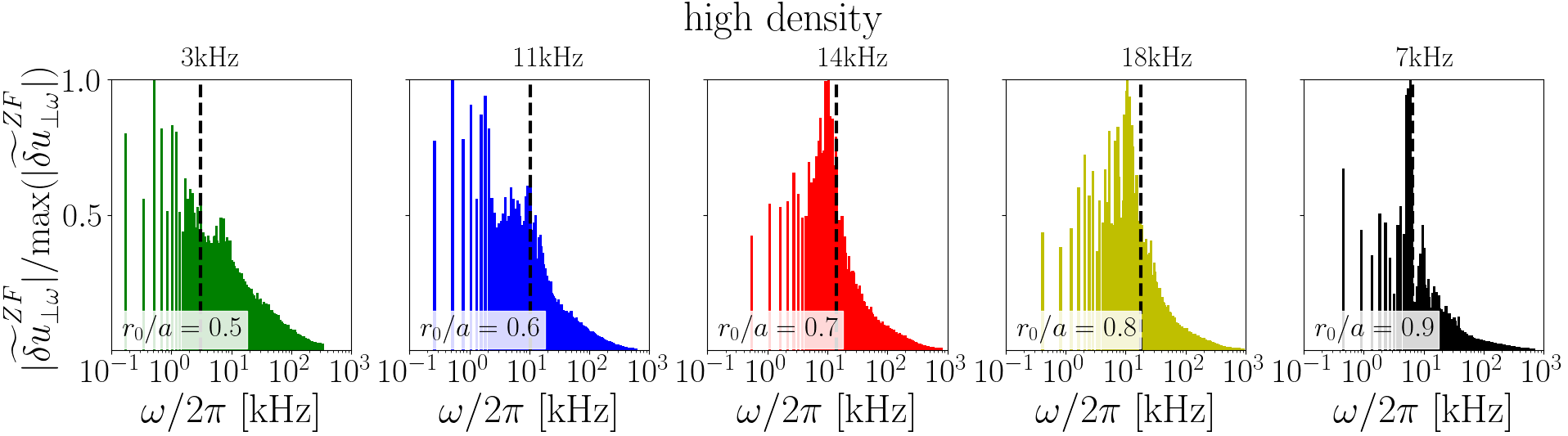}
    \caption{Normalized amplitudes $|\widetilde{\delta u}_{\perp \omega}^{\mathrm{ZF}}|$ computed with \texttt{stella} and represented as a function of the frequency $\omega$ for the low (top) and high (bottom) density scenarios. The dashed vertical lines indicate the frequency $\omega_0$ such that  $|\widetilde{\delta u}^{ZF}_{\perp\omega_0}| < \mathrm{max}(|\widetilde{\delta u}_{\perp\omega}^{ZF}|)/2$ for all $\omega > \omega_0$.}
    \label{fig:v_freq_ZF}
\end{figure}

\noindent
Taking a Fourier transform of $\delta u_{\perp}^{\mathrm{ZF}}(x,t)$ in $t$,  
\begin{equation}
    {\delta u}_{\perp}^{\mathrm{ZF}}(x,t) = \sum_{\omega}\widetilde{\delta u}_{\perp \omega}(x)\exp(-\mathrm{i} \omega t),
\end{equation}
allows us to obtain the frequency spectrum, $\widetilde{\delta u}_{\perp \omega}$, for each $x$ position of our flux tube. To improve the statistics of our results, we perform an average along $x$, defined for each frequency of the spectrum as 
\begin{equation}
   {\widetilde{\delta u}_{\perp \omega}^{\mathrm{ZF}}} = \frac{1}{N_x}{\sum_{j}\widetilde{\delta u}_{\perp \omega} (x_j)},
\end{equation}
where $N_x$ is the number of grid points along the $x$ direction. Figure \ref{fig:v_freq_ZF} shows the amplitudes of ${\widetilde{\delta u}_{\perp\omega}^{\mathrm{ZF}}}$ normalized to their maximum value, eliminating the mode $\omega=0$ (which corresponds to the time-averaged value). For the radial positions $r_0/a = \{0.7,0.8,0.9\}$ dominant frequencies cluster in the range $\omega/2\pi\sim [5,10]$ kHz. For the innermost positions, $r_0/a=\{0.5, 0.6\}$, dominant frequencies group around the lower boundary of $\omega$ with no clear peak in the spectra. In figure \ref{fig:v_freq_ZF}, we have also depicted with vertical dashed lines the frequency $\omega_{0}$ above which the amplitude $|{\widetilde{\delta u}_{\perp\omega}^{\mathrm{ZF}}}|$ is always smaller than half the maximum value in each case. It can be observed that, for all radial positions, frequencies $\omega_0/2\pi\leq 20$ kHz dominate the spectrum of the fluctuating zonal $E\times B$ flow. 

\section{Comparisons between linear and nonlinear calculations of $\varphi$}\label{sec:potential}
    The analysis presented in this work has focused, so far, on numerical quantities that can be directly compared against DR measurements. Nevertheless, aspects beyond that comparison, like the nature of the background turbulence, its localization along the flux tube or its linear properties have not been discussed. In the present section we briefly address some of these characteristics, comparing the linear and nonlinear frequency spectra of the electrostatic potential in order to assess to what extent linear frequencies remain during the nonlinear phase.\\

\begin{figure}[H]
    \centering
    \includegraphics[width=0.97\linewidth]{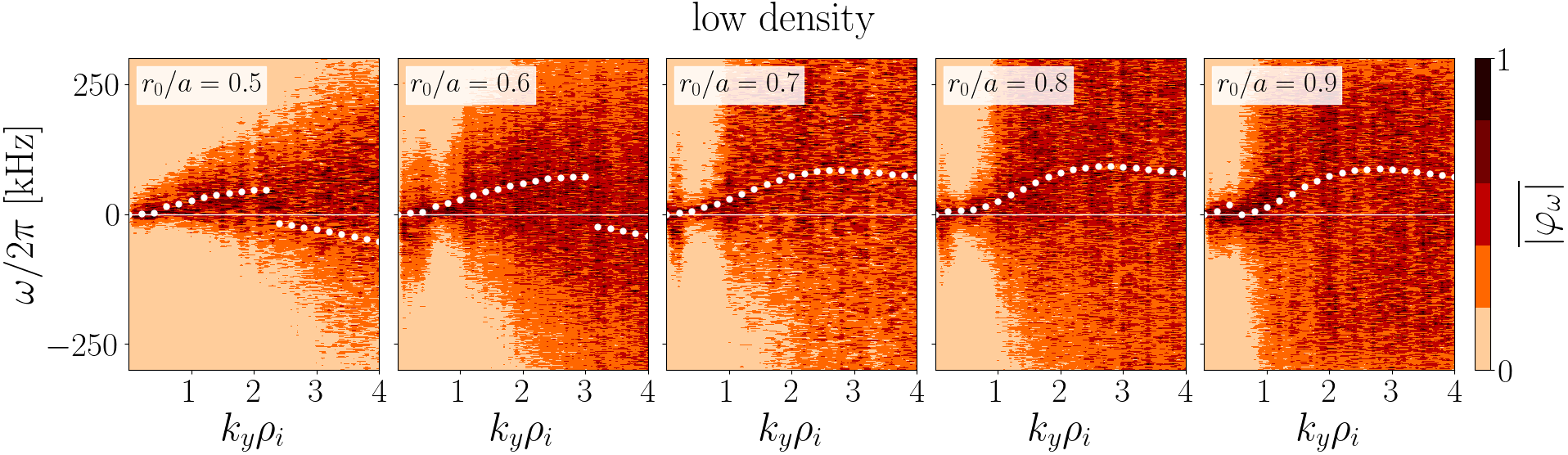}
    \includegraphics[width=0.97\linewidth]{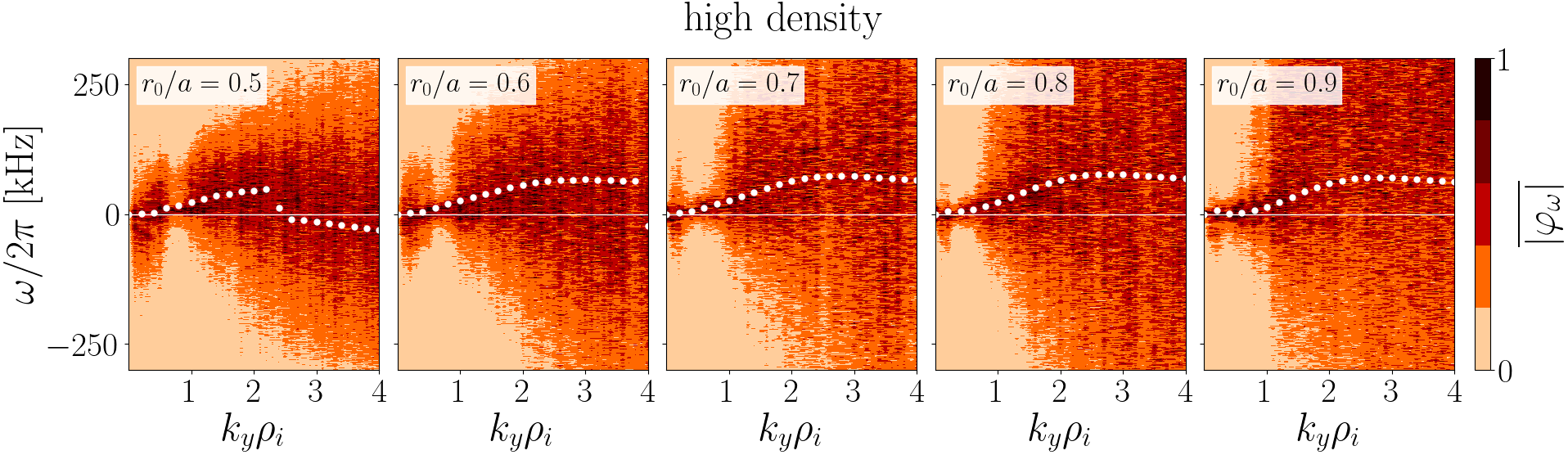}
    \caption{Nonlinear frequency spectrum of $\sum_{k_x}\langle\widehat{\varphi}_{k_x,k_y}\rangle_{\zeta}$ for the low (top) and high (bottom) density scenarios computed with \texttt{stella} as a function of $k_y$. The amplitudes $|\widetilde{\varphi}_{\omega}|$ are normalized to their maximum value at each $k_y$. The white dots are the results of linear $k_y$ scans assuming $k_x=0$.}
    \label{fig:phi_freq}
\end{figure}

\begin{figure}[H]
    \centering
    \includegraphics[width=0.97\linewidth]{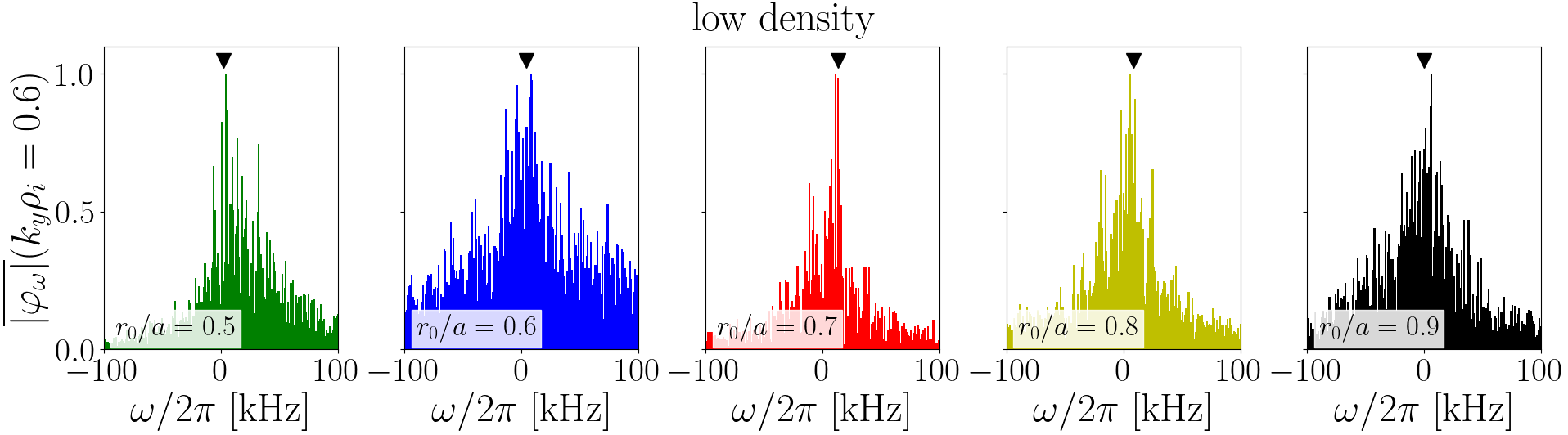}
    \includegraphics[width=0.97\linewidth]{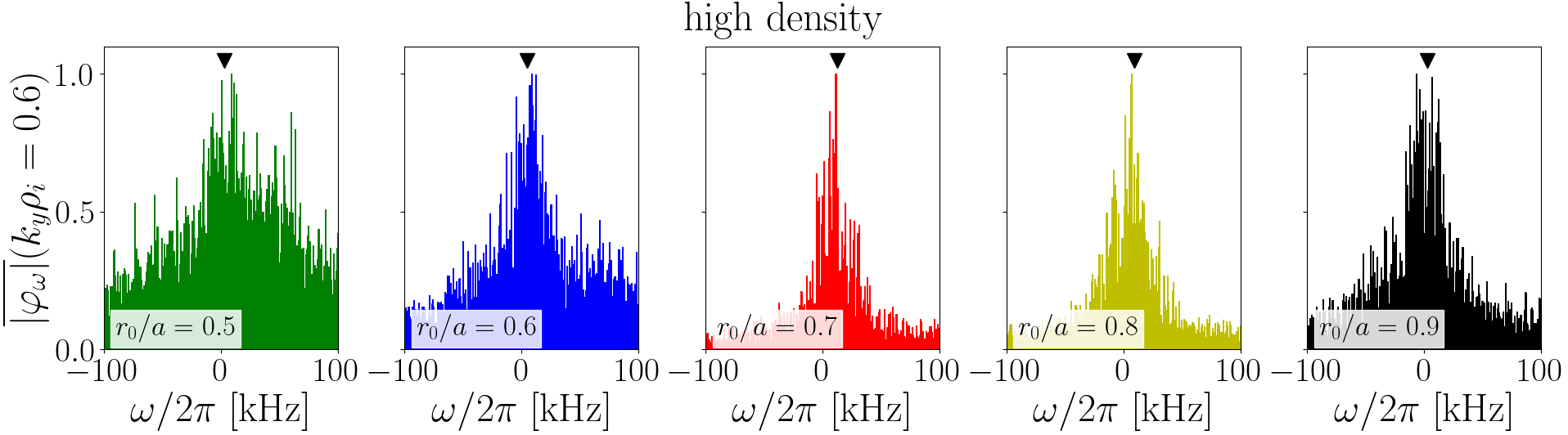}
    \caption{Nonlinear results of $\overline{|{\varphi}_{\omega}|}$ for the low (top) and high (bottom) density scenarios, evaluated at $k_y\rho_i=0.6$. The linear result for the mode $(k_x\rho_i, k_y\rho_i)=(0,0.6)$ is represented as a black triangle in every situation.}%
   \label{fig:low_k}
\end{figure}

In linear simulations, the time evolution of each $\left<\widehat{\varphi}_{k_x,k_y}\right>_{\zeta}(t)$ is assumed to be proportional to $\exp[(\gamma - \mathrm{i}\omega_r) t],$ with $\omega_r$ and $\gamma$ the linear frequency and growth rate. On the other hand, from the saturated phase obtained from nonlinear simulations, we can calculate the frequency spectrum for each $k_y$ mode of $\langle \varphi \rangle$, summing over all $k_x$ components as

\begin{equation}
    \sum_{k_x}\langle \widehat{\varphi}_{k_x,k_y} \rangle_{\zeta}(t) = \sum_{\omega} \widetilde{\varphi}_{\omega}(k_y)\exp(-\mathrm{i}\omega t).
\end{equation}
This Fourier decomposition allows a direct comparison between the nonlinear frequency spectrum of $\langle\varphi\rangle$ and the frequencies obtained from linear simulations. Figure \ref{fig:phi_freq} shows, as a function of $\omega$, the amplitude of $\widetilde{\varphi}_{\omega}$ normalized to the largest value found (at a certain $\omega = \omega^{M}$) for each $k_y$, i.e.

\begin{equation}
    \overline{|{\varphi}_{\omega}|}(k_y) = \frac{|\widetilde{{\varphi}}_{\omega}|(k_y)}{|\widetilde{{\varphi}}_{\omega=\omega^M}|(k_y)}.
\end{equation}
In addition, this figure also depicts the linear frequencies obtained from $k_y$ scans for a fixed $k_x=0$. Since the modes $\widehat{\varphi}_{k_x, k_y}$ are complex, we do not expect a symmetry in $|\widetilde{\varphi}_{\omega}|$ with respect to $\omega=0$ and, in contrast to section \ref{sec:predictions}, we have represented both positive and negative values of $\omega$ in figure \ref{fig:phi_freq}. It can be seen that at high $k_y$, the frequency spectra broadens and a dominant frequency in the nonlinear spectra is lacking. It is interesting though that for the wavenumbers measured by the DR system, changes in the sign of the linear frequency take place. Specifically, for the two analyzed scenarios, the linear frequency is negative for $r_0/a=0.5$ at $k_y=k_y^{\mathrm{DR}}$, whereas it is positive for all other radii, which points out to a change in the propagation direction of the drift waves driving the instability. On the other hand, for low $k_y$, the linear frequency values are reasonably close to those of the dominant nonlinear ones, which correspond with the darkest regions of the maps in figure \ref{fig:phi_freq}. This fact can be more clearly observed in figure \ref{fig:low_k}, where the normalized amplitudes $\overline{|{\varphi}_{\omega}|}(k_y)$ are represented as a function of $\omega$ for $k_y\rho_i = 0.6$. In that case, the linear frequency, indicated by a triangle in the figure, is located fairly close to the peak of the nonlinear frequency spectrum for every radial position analyzed. Similar agreements have been reported for the ASDEX Upgrade tokamak in \cite{Told_2013}, comparing linear and nonlinear simulations carried out with \texttt{GENE}.

\noindent

\begin{figure}[H]
    \centering
    \begin{subfigure}[b]{0.48\linewidth}        
        \centering
        \includegraphics[width=\linewidth]{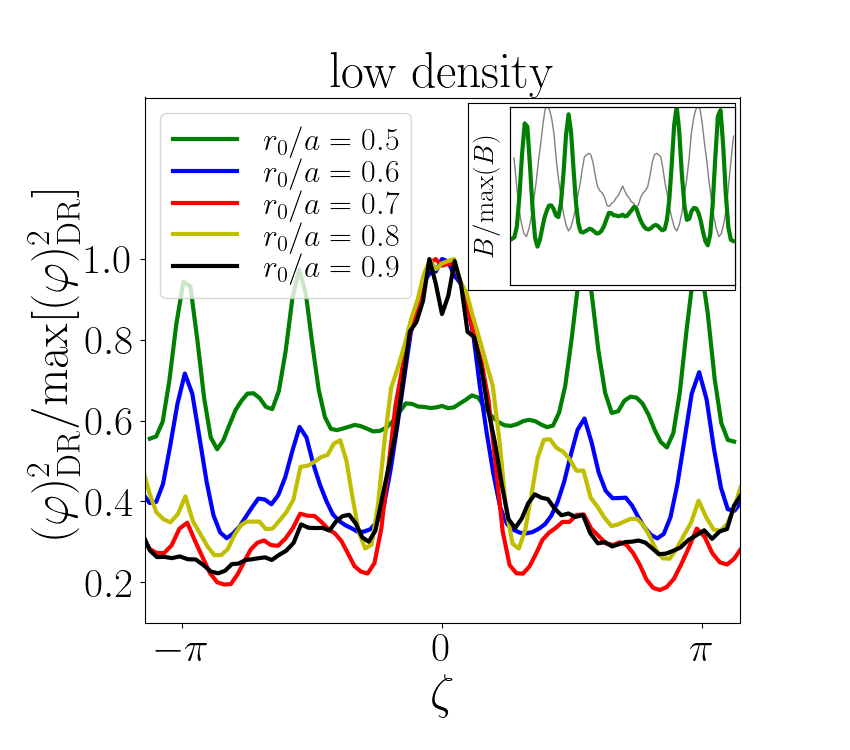}
    \end{subfigure}
\begin{subfigure}[b]{0.48\linewidth}        
        \centering
        \includegraphics[width=\linewidth]{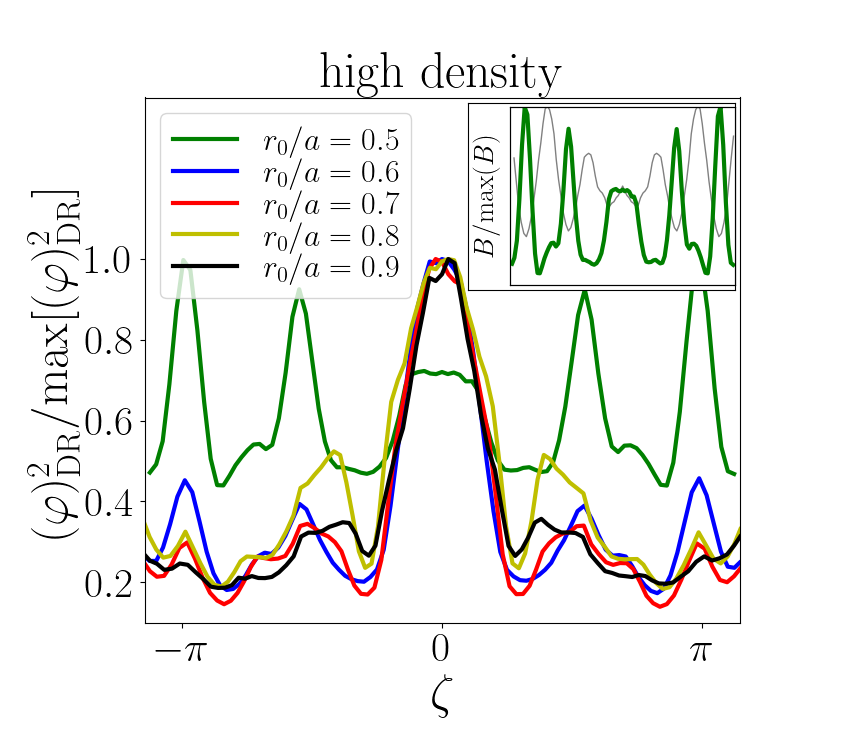}
    \end{subfigure}
    \caption{Normalized $(\varphi)^2_{\mathrm{DR}}$ as a function of $\zeta$, computed with \texttt{stella} for the low (left) and high (right) density scenarios. The insets show the results corresponding to $r_0/a=0.5$, together with the magnetic field strength as a gray solid line.}
    \label{fig:phi_z}
\end{figure}

\begin{figure}[H]
    \centering
    \begin{subfigure}[b]{0.48\linewidth}        
        \centering
        \includegraphics[width=\linewidth]{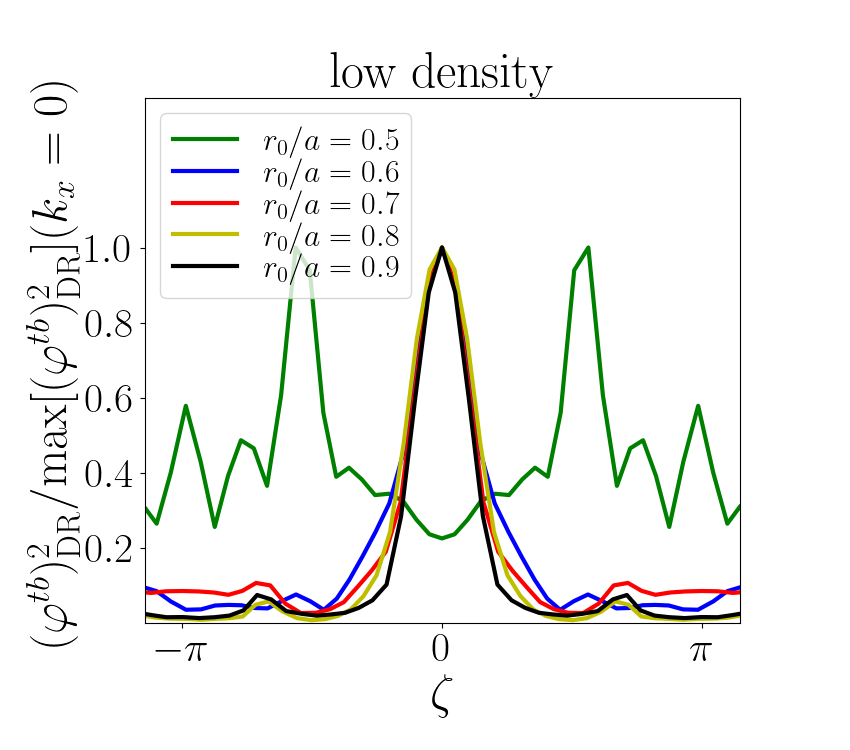}
    \end{subfigure}
\begin{subfigure}[b]{0.48\linewidth}        
        \centering
        \includegraphics[width=\linewidth]{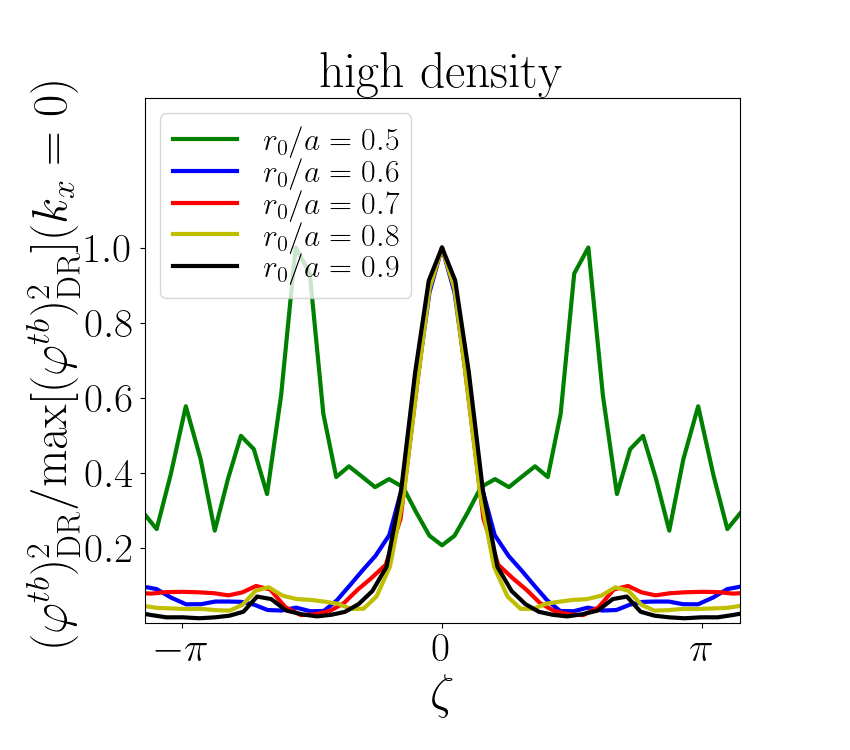}
    \end{subfigure}
    \caption{\red{Normalized $(\varphi)^2_{\mathrm{DR}}$ for the linearly unstable mode with $k_x=0$ as a function of $\zeta$, computed with \texttt{stella} for the low (left) and high (right) density scenarios.}}
    \label{fig:phi_z_linear}
\end{figure}

Finally, we analyze the localization of the fluctuating electrostatic potential along the flux tube for the wavenumber explored by the DR system. In particular, in figure \ref{fig:phi_z} we represent the time-averaged squared amplitude of $\varphi$ evaluated at $k_y=k_{y}^{\mathrm{DR}}$, this is
    \begin{equation}
        (\varphi)^2_{\mathrm{DR}} (\zeta) =\biggl< \sum_{k_x}\abs{\widehat{\varphi}_{k_x, k_y = k_{y}^{\mathrm{DR}}}}^2(\zeta,t)\biggr>_t. 
    \end{equation}
\red{To assess the correlation between the structure of nonlinear modes at $k_y^{\mathrm{DR}}$ and those computed for the most linearly unstable mode for the same $k_y^{DR}$ (generally located and obtained at $k_x=0$), we have depicted the parallel structure of the electrostatic potential from these linear simulations in figure \ref{fig:phi_z_linear}. It is observed that both linear and nonlinear results exhibit similar structures, with slight differences possibly due to the contribution of every $k_x$ mode considered in the nonlinear analysis and, of course, nonlinear effects. In particular, } $(\varphi)^2_{\mathrm{DR}}$ is strongly localized around the center of the flux tube, $\zeta = 0$, for every radial position except at $r_0/a=0.5$. That localization in $\zeta$ overlaps with regions of bad curvature in this configuration of W7-X. This suggests that ion-temperature-gradient (ITG) modes \cite{Biglari1989} play a leading role at $r_0/a>0.5$. {In addition, 
we can conclude that the linear modes at $k_y=k_y^{\mathrm{DR}}$ propagate in the ion diamagnetic direction, \red{which in the present work correspond with positive values of $\omega$} (see white dots in figure \ref{fig:phi_freq}),} as it is expected for ITG modes. On the other hand, the fluctuating electrostatic potential for the position $r_0/a=0.5$ is localized at regions of magnetic field wells (see the insets of figure \ref{fig:phi_z}, where the structure of this fluctuating electrostatic potential is represented together with the magnetic field strength). This fact, and the change in the sign of the linear frequency, points out to a prominent role of trapped electrons on the turbulence at $r_0/a=0.5$. In summary, looking at $k_y=k_y^{\mathrm{DR}}$, where the standard linear stability analysis finds a change in the sign of the frequency, a very different localization of the turbulent electrostatic potential is found nonlinearly.  

\section{Summary and conclusions}\label{sec:conclusions}
In the present work we have used the electrostatic flux tube version of the gyrokinetic code \texttt{stella} with the aim of calculating important turbulent quantities that can be directly compared with Doppler reflectometry measurements in the W7-X stellarator. Five radial positions of two ECRH discharges corresponding to the first operation phase (OP1) of W7-X have been considered throughout the paper.

In the first place, nonlinear simulations have been carried out to compute the amplitude of the density fluctuations at the spatial {region} and perpendicular wavenumbers explored by the DR system. Numerical results and DR measurements of the squared density fluctuations cover a range of about 15 dB and feature in both cases a monotonic increase with the radial coordinate. This is due to the different regions of the wavenumber spectrum accessed by the diagnostic at each radial location. Discrepancies between the numerical and experimental results in the different fluctuation levels between the two scenarios have also been observed. To expand the numerical characterization of the turbulence towards other quantities measurable by DR, the frequency spectra of the density fluctuations and of the zonal flow fluctuations have been provided. With regard to the density fluctuations, in general, the largest amplitudes are found for frequencies around a few tens of kHz. For higher frequencies, a slight decrease follows up to approximate $\omega/2\pi\simeq 500$ kHz, where an abrupt fall of the amplitude is found. Concerning the frequency spectra of the zonal flow fluctuations, they are dominated by frequencies with $\omega/2\pi\leq 20$ kHz and find their peak values for $\omega/2\pi\lesssim 10$ kHz at most radial positions. Future highly time-resolved characterization of the backscattered power and measurements of $u_{\perp}$ fluctuations at distant locations over the same flux surface will allow to validate these numerical findings. Finally, the turbulent electrostatic potential has been provided and compared against linear simulations, showing, for low $k_y$ values, that the dominant frequencies of the nonlinear spectrum cluster nearly around the frequency of the most unstable mode linearly. In addition, these results have also shown that, at $r_0/a=0.5$, trapped electrons seem to play a leading role on the background turbulence.

\appendix



\section{Poloidal profiles of $(\delta n)^2_{\mathrm{DR}}$}\label{ap:dens_poloidal}
\begin{figure}[H]
    \centering
    \begin{subfigure}[b]{0.48\linewidth}        
        \centering
        \includegraphics[width=\linewidth]{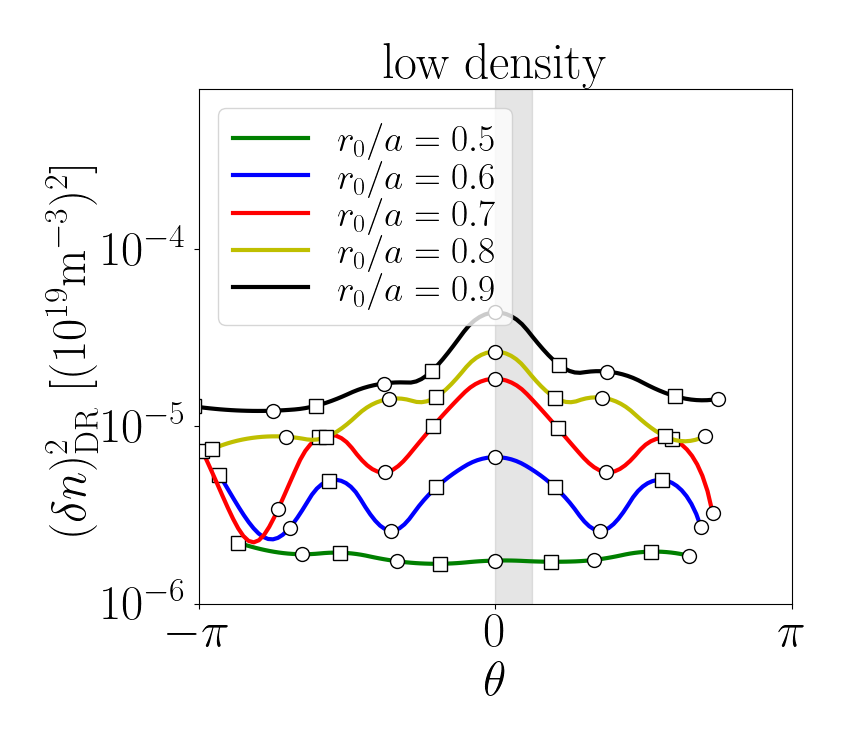}
    \end{subfigure}
\begin{subfigure}[b]{0.48\linewidth}        
        \centering
        \includegraphics[width=\linewidth]{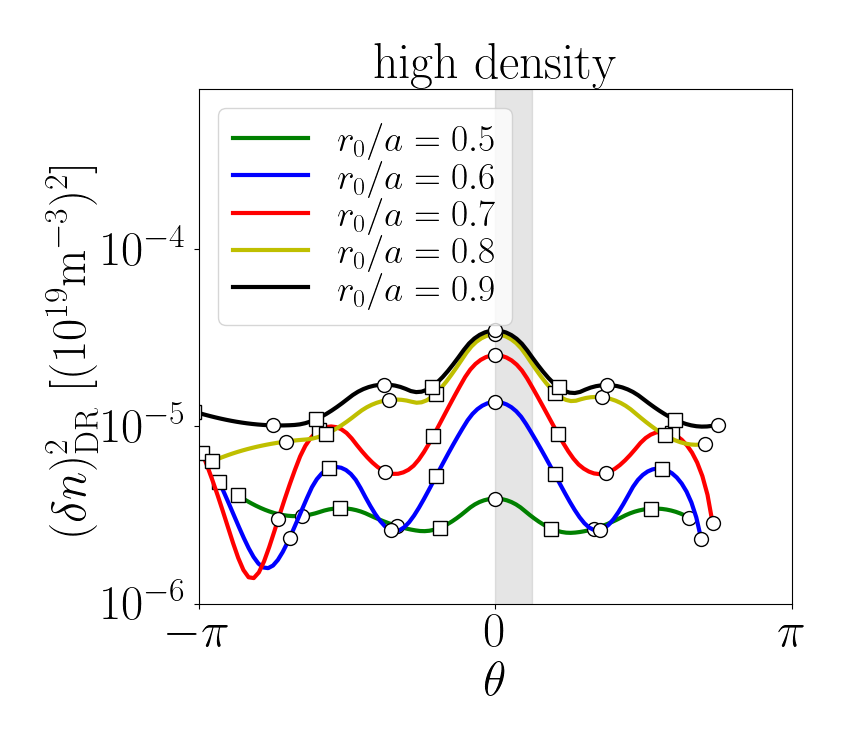}
    \end{subfigure}
    \caption{$(\delta n)^2_{\mathrm{DR}}$ as a function of the poloidal coordinate $\theta$ for a fixed $\zeta=0$ for the low (left) and high (right) density scenarios. Computed as an interpolation of the $(\delta n)^2_{\mathrm{DR}}$ calculated for the bean (circles) and triangular (squares) flux tubes. The shadowed region shows approximately the maximum deviation of the DR measurement position from $\theta=0$.}
    \label{fig:dens_theta}
\end{figure}

     The numerical study presented in this work has been performed considering the position $(\theta, \zeta)=(0,0)$, with respect to which the bean flux tube is centered. However, as it is shown in figure \ref{fig:DR_view}, that position does not correspond to the exact location where the beam launched by the DR system reflects and, consequently, where the measurement is performed. In order to assess the impact of this deviation in the results presented in section \ref{sec:comparisons}, we have obtained poloidal profiles of $(\delta n)_{\mathrm{DR}}^2$, that are represented in figure \ref{fig:dens_theta}. For this, $(\delta n)^2_{\mathrm{DR}}$ has been interpolated considering its value at the points of the plane $\zeta=0$ crossed by the bean and triangular flux tubes (see figure \ref{fig:DR_view}). The maximum poloidal deviation of the DR measurement points with respect to $\theta=0$, which is found at the innermost radii, is also indicated as a shadowed region in figure \ref{fig:dens_theta}. From these results, deviations $>20\%$ between $(\delta n)^2_{\mathrm{DR}}$ evaluated at the exact measurement location and that at $(\theta, \zeta)= (0,0)$ are not expected. Such differences are negligible compared to the quantitative discrepancy between the simulations and the experimental measurements shown in figure \ref{fig:dens_r_DR}. Hence, considering the position $(\theta, \zeta)=(0,0)$ is a good approximation to the real location where the DR system measures. 

\ack{
    AGJ and JMGR acknowledge useful discussions with Hanne Thienpondt on the development of the postprocessing tools. This work has been carried out within the framework of the EUROfusion Consortium, funded by the European Union via the Euratom Research and Training Programme (Grant Agreement No 101052200 - EUROfusion). Views and opinions expressed are however those of the author(s) only and do not necessarily reflect those of the European Union or the European Commission. Neither the European Union nor the European Commission can be held responsible for them. This research was supported in part by grant PID2021-123175NB-I00, Ministerio de Ciencia e Innovación, Spain. The simulations were carried out in the clusters Marconi (Cineca, Italy) and Xula (Ciemat, Spain).
    }


\section*{References}
\bibliographystyle{unsrt}
\bibliography{bib_PHD}

\begin{thebibliography}{10}

\bibitem{wolf2017}
R.~C. Wolf, A.~Ali, A.~Alonso, J.~Baldzuhn, C.~Beidler, M.~Beurskens,
  C.~Biedermann, H.~S. Bosch, S.~Bozhenkov, R.~Brakel, A.~Dinklage, Y.~Feng,
  G.~Fuchert, J.~Geiger, O.~Grulke, P.~Helander, M.~Hirsch, U.~H\"{o}fel,
  M.~Jakubowski, J.~Knauer, G.~Kocsis, R.~K\"{o}nig, P.~Kornejew,
  A.~Kr\"{a}mer-Flecken, M.~Krychowiak, M.~Landreman, A.~Langenberg, H.P.
  Laqua, S.~Lazerson, H.~Maa{\ss}berg, S.~Marsen, M.~Marushchenko, D.~Moseev,
  H.~Niemann, N.~Pablant, E.~Pasch, K.~Rahbarnia, G.~Schlisio, T.~Stange,
  T.~Sunn Pedersen, J.~Svensson, T.~Szepesi, H.~Trimino Mora, Y.~Turkin,
  T.~Wauters, G.~Weir, U.~Wenzel, T.~Windisch, G.~Wurden, D.~Zhang,
  I.~Abramovic, S.~\"{A}k\"{a}slompolo, P.~Aleynikov, K.~Aleynikova,
  R.~Alzbutas, G.~Anda, T.~Andreeva, E.~Ascasibar, J.~Assmann, S.~G. Baek,
  M.~Banduch, T.~Barbui, M.~Barlak, K.~Baumann, W.~Behr, A.~Benndorf,
  O.~Bertuch, W.~Biel, D.~Birus, B.~Blackwell, E.~Blanco, M.~Blatzheim,
  T.~Bluhm, D.~B\"{o}ckenhoff, P.~Bolgert, M.~Borchardt, V.~Borsuk, J.~Boscary,
  L.~G. B\"{o}ttger, H.~Brand, Ch. Brandt, T.~Br\"{a}uer, H.~Braune,
  S.~Brezinsek, K.~J. Brunner, B.~Br\"{u}nner, R.~Burhenn, B.~Buttensch\"{o}n,
  V.~Bykov, I.~Calvo, B.~Cannas, A.~Cappa, A.~Carls, L.~Carraro, B.~Carvalho,
  F.~Castejon, A.~Charl, F.~Chernyshev, M.~Cianciosa, R.~Citarella,
  {\L}.~Ciupi{\'{n}}ski, G.~Claps, M.~Cole, M.J. Cole, F.~Cordella, G.~Cseh,
  A.~Czarnecka, A.~Czermak, K.~Czerski, M.~Czerwinski, G.~Czymek, A.~da~Molin,
  A.~da~Silva, G.~Dammertz, J.~Danielson, A.~de~la Pena, S.~Degenkolbe,
  P.~Denner, D.~P. Dhard, M.~Dostal, M.~Drevlak, P.~Drewelow, Ph. Drews,
  A.~Dudek, G.~Dundulis, F.~Durodie, P.~van Eeten, F.~Effenberg, G.~Ehrke,
  M.~Endler, D.~Ennis, E.~Erckmann, H.~Esteban, T.~Estrada, N.~Fahrenkamp,
  J.~H. Feist, J.~Fellinger, H.~Fernandes, W.H. Fietz, W.~Figacz,
  J.~Fontdecaba, O.~Ford, T.~Fornal, H.~Frerichs, A.~Freund, M.~F\"{u}hrer,
  T.~Funaba, A.~Galkowski, G.~Gantenbein, Y.~Gao, J.~M.~Garc{\'{\i}}a
  Rega{\~{n}}a, M.~Garcia-Munoz, D.~Gates, G.~Gawlik, B.~Geiger, V.~Giannella,
  N.~Gierse, A.~Gogoleva, B.~Goncalves, A.~Goriaev, D.~Gradic, M.~Grahl,
  J.~Green, A.~Grosman, H.~Grote, M.~Gruca, C.~Guerard, L.~Haiduk, X.~Han,
  F.~Harberts, J.~H. Harris, H.~J. Hartfu{\ss}, D.~Hartmann, D.~Hathiramani,
  B.~Hein, B.~Heinemann, P.~Heitzenroeder, S.~Henneberg, C.~Hennig,
  J.~Hernandez Sanchez, C.~Hidalgo, H.~H\"{o}lbe, K.P. Hollfeld,
  A.~H\"{o}lting, D.~H\"{o}schen, M.~Houry, J.~Howard, X.~Huang, M.~Huber,
  V.~Huber, H.~Hunger, K.~Ida, T.~Ilkei, S.~Illy, B.~Israeli, A.~Ivanov,
  S.~Jablonski, J.~Jagielski, J.~Jelonnek, H.~Jenzsch, P.~Junghans,
  J.~Kacmarczyk, T.~Kaliatka, J.~P. Kallmeyer, U.~Kamionka, R.~Karalevicius,
  H.~Kasahara, W.~Kasparek, N.~Kenmochi, M.~Keunecke, A.~Khilchenko, D.~Kinna,
  R.~Kleiber, T.~Klinger, M.~Knaup, Th. Kobarg, F.~K\"{o}chl, Y.~Kolesnichenko,
  A.~K\"{o}nies, M.~K\"{o}ppen, J.~Koshurinov, R.~Koslowski, F.~K\"{o}ster,
  R.~Koziol, M.~Kr\"{a}mer, R.~Krampitz, P.~Kraszewsk, N.~Krawczyk,
  T.~Kremeyer, Th. Krings, J.~Krom, G.~Krzesinski, I.~Ksiazek, M.~Kubkowska,
  G.~K\"{u}hner, T.~Kurki-Suonio, S.~Kwak, R.~Lang, S.~Langish, H.~Laqua,
  R.~Laube, C.~Lechte, M.~Lennartz, W.~Leonhardt, L.~Lewerentz, Y.~Liang, Ch.
  Linsmeier, S.~Liu, J.-F. Lobsien, D.~Loesser, J.~Loizu Cisquella, J.~Lore,
  A.~Lorenz, M.~Losert, L.~Lubyako, A.~L\"{u}cke, A.~Lumsdaine, V.~Lutsenko,
  J.~Majano-Brown, O.~Marchuk, M.~Mardenfeld, P.~Marek, S.~Massidda,
  S.~Masuzaki, D.~Maurer, K.~McCarthy, P.~McNeely, A.~Meier, D.~Mellein,
  B.~Mendelevitch, Ph. Mertens, D.~Mikkelsen, O.~Mishchenko, B.~Missal,
  J.~Mittelstaedt, T.~Mizuuchi, A.~Mollen, V.~Moncada, T.~M\"{o}nnich,
  T.~Morizaki, R.~Munk, S.~Murakami, F.~Musielok, G.~N{\'{a}}fr{\'{a}}di,
  M.~Nagel, D.~Naujoks, H.~Neilson, O.~Neubauer, U.~Neuner, T.~Ngo,
  R.~Nocentini, C.~N\"{u}hrenberg, J.~N\"{u}hrenberg, S.~Obermayer,
  G.~Offermanns, K.~Ogawa, J.~Ongena, J.~W. Oosterbeek, G.~Orozco, M.~Otte,
  L.~Pacios Rodriguez, W.~Pan, N.~Panadero, N.~Panadero Alvarez, A.~Panin,
  D.~Papenfu{\ss}, S.~Paqay, A.~Pavone, E.~Pawelec, G.~Pelka, X.~Peng,
  V.~Perseo, B.~Peterson, A.~Pieper, D.~Pilopp, S.~Pingel, F.~Pisano, B.~Plaum,
  G.~Plunk, M.~Povilaitis, J.~Preinhaelter, J.~Proll, M.~E. Puiatti, A.~Puig
  Sitjes, F.~Purps, M.~Rack, S.~R{\'{e}}csei, A.~Reiman, D.~Reiter, F.~Remppel,
  S.~Renard, R.~Riedl, J.~Riemann, S.~Rimkevicius, K.~Ri{\ss}e, A.~Rodatos,
  H.~R\"{o}hlinger, M.~Rom{\'{e}}, P.~Rong, H.-J. Roscher, B.~Roth,
  L.~Rudischhauser, K.~Rummel, T.~Rummel, A.~Runov, N.~Rust, L.~Ryc,
  S.~Ryosuke, R.~Sakamoto, A.~Samartsev, M.~Sanchez, F.~Sano, S.~Satake,
  G.~Satheeswaran, J.~Schacht, F.~Schauer, T.~Scherer, A.~Schlaich, K.~H.
  Schl\"{u}ter, J.~Schmitt, H.~Schmitz, O.~Schmitz, S.~Schmuck, M.~Schneider,
  W.~Schneider, M.~Scholz, P.~Scholz, R.~Schrittwieser, M.~Schr\"{o}der,
  T.~Schr\"{o}der, R.~Schroeder, H.~Schumacher, B.~Schweer, B.~Shanahan, I.~V.
  Shikhovtsev, M.~Sibilia, P.~Sinha, S.~Sipli\"{a}, J.~Skodzik, C.~Slaby,
  H.~Smith, W.~Spiess, D.A. Spong, A.~Spring, R.~Stadler, B.~Standley,
  L.~Stephey, M.~Stoneking, U.~Stridde, Z.~Sulek, C.~Surko, Y.~Suzuki,
  V.~Szab{\'{o}}, T.~Szabolics, Z.~Sz\"{o}kefalvi-Nagy, N.~Tamura, A.~Terra,
  J.~Terry, J.~Thomas, H.~Thomsen, M.~Thumm, C.P. von Thun, D.~Timmermann,
  P.~Titus, K.~Toi, J.M. Travere, P.~Traverso, J.~Tretter, H.~Tsuchiya,
  T.~Tsujimura, S.~Tulip{\'{a}}n, M.~Turnyanskiy, B.~Unterberg, J.~Urban,
  E.~Urbonavicius, I.~Vakulchyk, S.~Valet, B.~van Millingen, L.~Vela, J.~L.
  Velasco, M.~Vergote, M.~Vervier, N.~Vianello, H.~Viebke, R.~Vilbrandt,
  A.~Vork\"{o}rper, S.~Wadle, F.~Wagner, E.~Wang, N.~Wang, F.~Warmer,
  L.~Wegener, J.~Weggen, Y.~Wei, J.~Wendorf, A.~Werner, B.~Wiegel, F.~Wilde,
  E.~Winkler, V.~Winters, S.~Wolf, J.~Wolowski, A.~Wright, P.~Xanthopoulos,
  H.~Yamada, I.~Yamada, R.~Yasuhara, M.~Yokoyama, J.~Zajac, M.~Zarnstorff,
  A.~Zeitler, H.~Zhang, J.~Zhu, M.~Zilker, A.~Zimbal, A.~Zocco, S.~Zoletnik,
  and M.~Zuin.
\newblock Major results from the first plasma campaign of the {W}endelstein
  7-{X} stellarator.
\newblock {\em Nuclear Fusion}, 57(10):102020, July 2017.

\bibitem{Klinger2019}
T.~Klinger, T.~Andreeva, S.~Bozhenkov, C.~Brandt, R.~Burhenn,
  B.~Buttensch\"{o}n, G.~Fuchert, B.~Geiger, O.~Grulke, H.P. Laqua, N.~Pablant,
  K.~Rahbarnia, T.~Stange, A.~von Stechow, N.~Tamura, H.~Thomsen, Y.~Turkin,
  T.~Wegner, I.~Abramovic, S.~\"{A}k\"{a}slompolo, J.~Alcuson, P.~Aleynikov,
  K.~Aleynikova, A.~Ali, A.~Alonso, G.~Anda, E.~Ascasibar, J.~P. B\"{a}hner,
  S.G. Baek, M.~Balden, J.~Baldzuhn, M.~Banduch, T.~Barbui, W.~Behr,
  C.~Beidler, A.~Benndorf, C.~Biedermann, W.~Biel, B.~Blackwell, E.~Blanco,
  M.~Blatzheim, S.~Ballinger, T.~Bluhm, D.~B\"{o}ckenhoff, B.~B\"{o}swirth,
  L.~G. B\"{o}ttger, M.~Borchardt, V.~Borsuk, J.~Boscary, H.~S. Bosch,
  M.~Beurskens, R.~Brakel, H.~Brand, T.~Br\"{a}uer, H.~Braune, S.~Brezinsek,
  K.~J. Brunner, R.~Bussiahn, V.~Bykov, J.~Cai, I.~Calvo, B.~Cannas, A.~Cappa,
  A.~Carls, D.~Carralero, L.~Carraro, B.~Carvalho, F.~Castejon, A.~Charl,
  N.~Chaudhary, D.~Chauvin, F.~Chernyshev, M.~Cianciosa, R.~Citarella,
  G.~Claps, J.~Coenen, M.~Cole, M.~J. Cole, F.~Cordella, G.~Cseh, A.~Czarnecka,
  K.~Czerski, M.~Czerwinski, G.~Czymek, A.~da~Molin, A.~da~Silva, H.~Damm,
  A.~de~la Pena, S.~Degenkolbe, C.~P. Dhard, M.~Dibon, A.~Dinklage, T.~Dittmar,
  M.~Drevlak, P.~Drewelow, P.~Drews, F.~Durodie, E.~Edlund, P.~van Eeten,
  F.~Effenberg, G.~Ehrke, S.~Elgeti, M.~Endler, D.~Ennis, H.~Esteban,
  T.~Estrada, J.~Fellinger, Y.~Feng, E.~Flom, H.~Fernandes, W.~H. Fietz,
  W.~Figacz, J.~Fontdecaba, O.~Ford, T.~Fornal, H.~Frerichs, A.~Freund,
  T.~Funaba, A.~Galkowski, G.~Gantenbein, Y.~Gao, J.~M.~Garc{\'{\i}}a
  Rega{\~{n}}a, D.~Gates, J.~Geiger, V.~Giannella, A.~Gogoleva, B.~Goncalves,
  A.~Goriaev, D.~Gradic, M.~Grahl, J.~Green, H.~Greuner, A.~Grosman, H.~Grote,
  M.~Gruca, C.~Guerard, P.~Hacker, X.~Han, J.~H. Harris, D.~Hartmann,
  D.~Hathiramani, B.~Hein, B.~Heinemann, P.~Helander, S.~Henneberg, M.~Henkel,
  J.~Hernandez Sanchez, C.~Hidalgo, M.~Hirsch, K.~P. Hollfeld, U.~H\"{o}fel,
  A.~H\"{o}lting, D.~H\"{o}schen, M.~Houry, J.~Howard, X.~Huang, Z.~Huang,
  M.~Hubeny, M.~Huber, H.~Hunger, K.~Ida, T.~Ilkei, S.~Illy, B.~Israeli,
  S.~Jablonski, M.~Jakubowski, J.~Jelonnek, H.~Jenzsch, T.~Jesche, M.~Jia,
  P.~Junghanns, J.~Kacmarczyk, J.~P. Kallmeyer, U.~Kamionka, H.~Kasahara,
  W.~Kasparek, Y.~O. Kazakov, N.~Kenmochi, C.~Killer, A.~Kirschner, R.~Kleiber,
  J.~Knauer, M.~Knaup, A.~Knieps, T.~Kobarg, G.~Kocsis, F.~K\"{o}chl,
  Y.~Kolesnichenko, A.~K\"{o}nies, R.~K\"{o}nig, P.~Kornejew, J.~P. Koschinsky,
  F.~K\"{o}ster, M.~Kr\"{a}mer, R.~Krampitz, A.~Kr\"{a}mer-Flecken,
  N.~Krawczyk, T.~Kremeyer, J.~Krom, M.~Krychowiak, I.~Ksiazek, M.~Kubkowska,
  G.~K\"{u}hner, T.~Kurki-Suonio, P.A. Kurz, S.~Kwak, M.~Landreman, P.~Lang,
  R.~Lang, A.~Langenberg, S.~Langish, H.~Laqua, R.~Laube, S.~Lazerson,
  C.~Lechte, M.~Lennartz, W.~Leonhardt, C.~Li, C.~Li, Y.~Li, Y.~Liang,
  C.~Linsmeier, S.~Liu, J.-F. Lobsien, D.~Loesser, J.~Loizu Cisquella, J.~Lore,
  A.~Lorenz, M.~Losert, A.~L\"{u}cke, A.~Lumsdaine, V.~Lutsenko,
  H.~Maa{\ss}berg, O.~Marchuk, J.H. Matthew, S.~Marsen, M.~Marushchenko,
  S.~Masuzaki, D.~Maurer, M.~Mayer, K.~McCarthy, P.~McNeely, A.~Meier,
  D.~Mellein, B.~Mendelevitch, P.~Mertens, D.~Mikkelsen, A.~Mishchenko,
  B.~Missal, J.~Mittelstaedt, T.~Mizuuchi, A.~Mollen, V.~Moncada,
  T.~M\"{o}nnich, T.~Morisaki, D.~Moseev, S.~Murakami, G.~N{\'{a}}fr{\'{a}}di,
  M.~Nagel, D.~Naujoks, H.~Neilson, R.~Neu, O.~Neubauer, U.~Neuner, T.~Ngo,
  D.~Nicolai, S.~K. Nielsen, H.~Niemann, T.~Nishizawa, R.~Nocentini,
  C.~N\"{u}hrenberg, J.~N\"{u}hrenberg, S.~Obermayer, G.~Offermanns, K.~Ogawa,
  J.~\"{O}lmanns, J.~Ongena, J.~W. Oosterbeek, G.~Orozco, M.~Otte, L.~Pacios
  Rodriguez, N.~Panadero, N.~Panadero Alvarez, D.~Papenfu{\ss}, S.~Paqay,
  E.~Pasch, A.~Pavone, E.~Pawelec, T.~S. Pedersen, G.~Pelka, V.~Perseo,
  B.~Peterson, D.~Pilopp, S.~Pingel, F.~Pisano, B.~Plaum, G.~Plunk,
  P.~P\"{o}l\"{o}skei, M.~Porkolab, J.~Proll, M.~E. Puiatti, A.~Puig Sitjes,
  F.~Purps, M.~Rack, S.~R{\'{e}}csei, A.~Reiman, F.~Reimold, D.~Reiter,
  F.~Remppel, S.~Renard, R.~Riedl, J.~Riemann, K.~Risse, V.~Rohde,
  H.~R\"{o}hlinger, M.~Rom{\'{e}}, D.~Rondeshagen, P.~Rong, B.~Roth,
  L.~Rudischhauser, K.~Rummel, T.~Rummel, A.~Runov, N.~Rust, L.~Ryc,
  S.~Ryosuke, R.~Sakamoto, M.~Salewski, A.~Samartsev, E.~Sanchez, F.~Sano,
  S.~Satake, J.~Schacht, G.~Satheeswaran, F.~Schauer, T.~Scherer, J.~Schilling,
  A.~Schlaich, G.~Schlisio, F.~Schluck, K.~H. Schl\"{u}ter, J.~Schmitt,
  H.~Schmitz, O.~Schmitz, S.~Schmuck, M.~Schneider, W.~Schneider, P.~Scholz,
  R.~Schrittwieser, M.~Schr\"{o}der, T.~Schr\"{o}der, R.~Schroeder,
  H.~Schumacher, B.~Schweer, E.~Scott, S.~Sereda, B.~Shanahan, M.~Sibilia,
  P.~Sinha, S.~Sipli\"{a}, C.~Slaby, M.~Sleczka, H.~Smith, W.~Spiess, D.A.
  Spong, A.~Spring, R.~Stadler, M.~Stejner, L.~Stephey, U.~Stridde, C.~Suzuki,
  J.~Svensson, V.~Szab{\'{o}}, T.~Szabolics, T.~Szepesi,
  Z.~Sz\"{o}kefalvi-Nagy, A.~Tancetti, J.~Terry, J.~Thomas, M.~Thumm, J.~M.
  Travere, P.~Traverso, J.~Tretter, H.~Trimino Mora, H.~Tsuchiya, T.~Tsujimura,
  S.~Tulip{\'{a}}n, B.~Unterberg, I.~Vakulchyk, S.~Valet, L.~Vano, B.~van
  Milligen, A.~J. van Vuuren, L.~Vela, J.~L. Velasco, M.~Vergote, M.~Vervier,
  N.~Vianello, H.~Viebke, R.~Vilbrandt, A.~Vork\"{o}per, S.~Wadle, F.~Wagner,
  E.~Wang, N.~Wang, Z.~Wang, F.~Warmer, T.~Wauters, L.~Wegener, J.~Weggen,
  Y.~Wei, G.~Weir, J.~Wendorf, U.~Wenzel, A.~Werner, A.~White, B.~Wiegel,
  F.~Wilde, T.~Windisch, M.~Winkler, A.~Winter, V.~Winters, S.~Wolf, R.~C.
  Wolf, A.~Wright, G.~Wurden, P.~Xanthopoulos, H.~Yamada, I.~Yamada,
  R.~Yasuhara, M.~Yokoyama, M.~Zanini, M.~Zarnstorff, A.~Zeitler, D.~Zhang,
  H.~Zhang, J.~Zhu, M.~Zilker, A.~Zocco, S.~Zoletnik, and M.~Zuin.
\newblock Overview of first {W}endelstein 7-{X} high-performance operation.
\newblock {\em Nuclear Fusion}, 59(11):112004, June 2019.

\bibitem{Bozhenkov2020}
S.~A. Bozhenkov, Y.~Kazakov, O.~P. Ford, M.~N.~A. Beurskens, J.~Alcus{\'{o}}n,
  J.~A. Alonso, J.~Baldzuhn, C.~Brandt, K.~J. Brunner, H.~Damm, G.~Fuchert,
  J.~Geiger, O.~Grulke, M.~Hirsch, U.~H\"{o}fel, Z.~Huang, J.~Knauer,
  M.~Krychowiak, A.~Langenberg, H.~P. Laqua, S.~Lazerson, N.~B. Marushchenko,
  D.~Moseev, M.~Otte, N.~Pablant, E.~Pasch, A.~Pavone, J.~H.~E. Proll,
  K.~Rahbarnia, E.~R. Scott, H.~M. Smith, T.~Stange, A.~von Stechow,
  H.~Thomsen, Yu. Turkin, G.~Wurden, P.~Xanthopoulos, D.~Zhang, and R.~C. Wolf.
\newblock High-performance plasmas after pellet injections in {W}endelstein
  7-{X}.
\newblock {\em Nuclear Fusion}, 60(6):066011, May 2020.

\bibitem{GarcaRegaa2021}
J.~M. Garc{\'{\i}}a-Rega{\~{n}}a, M.~Barnes, I.~Calvo, F.~I. Parra, J.~A.
  Alcus{\'{o}}n, R.~Davies, A.~Gonz{\'{a}}lez-Jerez, A.~Moll{\'{e}}n,
  E.~S{\'{a}}nchez, J.~L. Velasco, and A.~Zocco.
\newblock Turbulent impurity transport simulations in {W}endelstein 7-{X}
  plasmas.
\newblock {\em Journal of Plasma Physics}, 87(1):85587010, February 2021.

\bibitem{Geiger2019}
B.~Geiger, Th. Wegner, C.~D. Beidler, R.~Burhenn, B.~Buttensch\"{o}n, R.~Dux,
  A.~Langenberg, N.~A. Pablant, T.~P\"{u}tterich, Y.~Turkin, T.~Windisch,
  V.~Winters, M.~Beurskens, C.~Biedermann, K.~J. Brunner, G.~Cseh, H.~Damm,
  F.~Effenberg, G.~Fuchert, O.~Grulke, J.~H. Harris, C.~Killer, J.~Knauer,
  G.~Kocsis, A.~Kr\"{a}mer-Flecken, T.~Kremeyer, M.~Krychowiak, O.~Marchuk,
  D.~Nicolai, K.~Rahbarnia, G.~Satheeswaran, J.~Schilling, O.~Schmitz,
  T.~Schr\"{o}der, T.~Szepesi, H.~Thomsen, H.~Trimino Mora, P.~Traverso, and
  D.~Zhang.
\newblock Observation of anomalous impurity transport during low-density
  experiments in {W}7-{X} with laser blow-off injections of iron.
\newblock {\em Nuclear Fusion}, 59(4):046009, February 2019.

\bibitem{Maasberg1999}
H.~Maa{\ss}berg, C.~D. Beidler, and E.~E. Simmet.
\newblock Density control problems in large stellarators with neoclassical
  transport.
\newblock {\em Plasma Physics and Controlled Fusion}, 41(9):1135--1153, August
  1999.

\bibitem{Thienpondt2023}
H.~Thienpondt, J.~M. Garc{\'{\i}}a-Rega{\~{n}}a, I.~Calvo, J.~A. Alonso, J.~L.
  Velasco, A.~Gonz{\'{a}}lez-Jerez, M.~Barnes, K.~Brunner, O.~Ford, G.~Fuchert,
  J.~Knauer, E.~Pasch, and L.~Van{\'{o}} and.
\newblock Prevention of core particle depletion in stellarators by turbulence.
\newblock {\em Physical Review Research}, 5(2), June 2023.

\bibitem{Barnes2019}
M.~Barnes, F.~I. Parra, and M.~Landreman.
\newblock stella: An operator-split, implicit{\textendash}explicit
  $\delta$f-gyrokinetic code for general magnetic field configurations.
\newblock {\em Journal of Computational Physics}, 391:365--380, August 2019.

\bibitem{Maurer2020}
M.~Maurer, A.~Ba{\~{n}}{\'{o}}n-Navarro, T.~Dannert, M.~Restelli,
  F.~Hindenlang, T.~G\"{o}rler, D.~Told, D.~Jarema, G.~Merlo, and F.~Jenko.
\newblock {GENE}-3{D}: a global gyrokinetic turbulence code for stellarators.
\newblock {\em Journal of Computational Physics}, 420:109694, November 2020.

\bibitem{Mandell2023}
N.~R. Mandell, W.~Dorland, I.~Abel, R.~Gaur, P.~Kim, M.~Martin, and T.~Qian.
\newblock {GX}: a {GPU}-native gyrokinetic turbulence code for tokamak and
  stellarator design, 2022.

\bibitem{GonzlezJerez2022}
A.~Gonz{\'{a}}lez-Jerez, P.~Xanthopoulos, J.~M. Garc{\'{\i}}a-Rega{\~{n}}a,
  I.~Calvo, J.~Alcus{\'{o}}n, A.~Ba{\~{n}}{\'{o}}n Navarro, M.~Barnes, F.~I.
  Parra, and J.~Geiger.
\newblock Electrostatic gyrokinetic simulations in {W}endelstein 7-{X}
  geometry: benchmark between the codes stella and {GENE}.
\newblock {\em Journal of Plasma Physics}, 88(3), June 2022.

\bibitem{Sanchez2021}
E.~S{\'{a}}nchez, J.~M. Garc{\'{\i}}a-Rega{\~{n}}a, A.~Ba{\~{n}}{\'{o}}n
  Navarro, J.~H.~E. Proll, C.~Mora Moreno, A.~Gonz{\'{a}}lez-Jerez, I.~Calvo,
  R.~Kleiber, J.~Riemann, J.~Smoniewski, M.~Barnes, and F.~I. Parra.
\newblock Gyrokinetic simulations in stellarators using different computational
  domains.
\newblock {\em Nuclear Fusion}, 61(11):116074, October 2021.

\bibitem{Sanchez2023}
E.~S{\'{a}}nchez, A.~Ba{\~{n}}{\'{o}}n Navarro, F.~Wilms, M.~Borchardt,
  R.~Kleiber, and F.~Jenko.
\newblock Instabilities and turbulence in stellarators from the perspective of
  global codes.
\newblock {\em Nuclear Fusion}, 63(4):046013, March 2023.

\bibitem{GarcaRegaa2021NF}
J.~M. Garc{\'{\i}}a-Rega{\~{n}}a, M.~Barnes, I.~Calvo, A.~Gonz{\'{a}}lez-Jerez,
  H.~Thienpondt, E.~S{\'{a}}nchez, F.~I. Parra, and D.~A. St.-Onge.
\newblock Turbulent transport of impurities in 3{D} devices.
\newblock {\em Nuclear Fusion}, 61(11):116019, September 2021.

\bibitem{ThienpondtISHW2022}
H.~Thienpondt, J.~M. García-Regaña, I.~Calvo, A.~González-Jerez,
  M.~Barnesand~F. Parra, and R.~Davies.
\newblock Turbulent transport versus density gradient:an inter-machine study
  with the gyrokinetic code stella.
\newblock In {\em 23rd International Stellarator-Heliotron Workshop}, 2022.

\bibitem{Bhner2021}
J.~P. B\"{a}hner, J.~A. Alcus{\'{o}}n, S.~K. Hansen, A.~von Stechow, O.~Grulke,
  T.~Windisch, H.~M. Smith, Z.~Huang, E.~M. Edlund, M.~Porkolab, M.~N.~A.
  Beurskens, S.~A. Bozhenkov, O.~P. Ford, L.~Van{\'{o}}, A.~Langenberg,
  N.~Pablant, G.~G. Plunk, A.~Ba{\~{n}}{\'{o}}n Navarro, and F.~Jenko.
\newblock Phase contrast imaging measurements and numerical simulations of
  turbulent density fluctuations in gas-fuelled {ECRH} discharges in
  {W}endelstein 7-{X}.
\newblock {\em Journal of Plasma Physics}, 87(3), June 2021.

\bibitem{Hansen2022}
S.~K. Hansen, M.~Porkolab, J.~P. B\"{a}hner, Z.~Huang, A.~von Stechow,
  O.~Grulke, E.~M. Edlund, F.~Wilms, A.~Ba{\~{n}}{\'{o}}n Navarro, F.~Jenko,
  and E.~S{\'{a}}nchez.
\newblock Development of a synthetic phase contrast imaging diagnostic for
  turbulence studies at {W}endelstein 7-{X}.
\newblock {\em Plasma Physics and Controlled Fusion}, 64(9):095011, July 2022.

\bibitem{Carralero2021}
D.~Carralero, T.~Estrada, E.~Maragkoudakis, T.~Windisch, J.~A. Alonso,
  M.~Beurskens, S.~Bozhenkov, I.~Calvo, H.~Damm, O.~Ford, G.~Fuchert, J.~M.
  Garc{\'{\i}}a-Rega{\~{n}}a, N.~Pablant, E.~S{\'{a}}nchez, E.~Pasch, J.~L.
  Velasco, and the~{W}endelstein 7-{X}~team.
\newblock An experimental characterization of core turbulence regimes in
  {W}endelstein 7-{X}.
\newblock {\em Nuclear Fusion}, 61(9):096015, August 2021.

\bibitem{Windisch2018}
T.~Windisch, S.~Wolf, G.~M. Weir, S.~A. Bozhenkov, H.~Damm, G.~Fuchert,
  O.~Grulke, M.~Hirsch, W.~Kasparek, T.~Klinger, C.~Lechte, E.~Pasch, B.~Plaum,
  and E.~A.~Scott and.
\newblock Phased array doppler reflectometry at {W}endelstein 7-{X}.
\newblock {\em Review of Scientific Instruments}, 89(10):10H115, October 2018.

\bibitem{Estrada2021}
T.~Estrada, D.~Carralero, T.~Windisch, E.~S{\'{a}}nchez, J.~M.
  Garc{\'{\i}}a-Rega{\~{n}}a, J.~Mart{\'{\i}}nez-Fern{\'{a}}ndez, A.~de~la
  Pe{\~{n}}a, J.~L. Velasco, J.~A. Alonso, M.~Beurskens, S.~Bozhenkov, H.~Damm,
  G.~Fuchert, R.~Kleiber, N.~Pablant, E.~Pasch, and the {W}7-{X}~team.
\newblock Radial electric field and density fluctuations measured by doppler
  reflectometry during the post-pellet enhanced confinement phase in {W}7-{X}.
\newblock {\em Nuclear Fusion}, 61(4):046008, March 2021.

\bibitem{Carralero2022}
D.~Carralero, T.~Estrada, E.~Maragkoudakis, T.~Windisch, J.~A. Alonso, J.~L.
  Velasco, O.~Ford, M.~Jakubowski, S.~Lazerson, M.~Beurskens, S.~Bozhenkov,
  I.~Calvo, H.~Damm, G.~Fuchert, J.~M. Garc{\'{\i}}a-Rega{\~{n}}a,
  U.~H\"{o}fel, N.~Marushchenko, N.~Pablant, E.~S{\'{a}}nchez, H.~M. Smith,
  E.~Pasch, and T.~Stange.
\newblock On the role of density fluctuations in the core turbulent transport
  of {W}endelstein 7-{X}.
\newblock {\em Plasma Physics and Controlled Fusion}, 64(4):044006, February
  2022.

\bibitem{Catto1978}
P.~J. Catto.
\newblock Linearized gyro-kinetics.
\newblock {\em Plasma Physics}, 20(7):719--722, July 1978.

\bibitem{Beer1995}
M.~A. Beer, S.~C. Cowley, and G.~W. Hammett.
\newblock Field-aligned coordinates for nonlinear simulations of tokamak
  turbulence.
\newblock {\em Physics of Plasmas}, 2(7):2687--2700, August 1995.

\bibitem{Martin2018}
M.~F. Martin, M.~Landreman, P.~Xanthopoulos, N.~R. Mandell, and W.~Dorland.
\newblock The parallel boundary condition for turbulence simulations in low
  magnetic shear devices.
\newblock {\em Plasma Physics and Controlled Fusion}, 60(9):095008, July 2018.

\bibitem{Grimm1983}
R.~C. Grimm, R.~L. Dewar, and J.~Manickam.
\newblock Ideal {MHD} stability calculations in axisymmetric toroidal
  coordinate systems.
\newblock {\em Journal of Computational Physics}, 49(1):94--117, January 1983.

\bibitem{Hirshman_1986}
S.~P. Hirshman, K.~C. Shaing, W.~I. van Rij, C.~O. Beasley, and E.~C. Crume.
\newblock Plasma transport coefficients for nonsymmetric toroidal confinement
  systems.
\newblock {\em Physics of Fluids}, 29(9):2951--2959, sep 1986.

\bibitem{Geiger2015}
J.~Geiger, C.~D. Beidler, Y.~Feng, H.~Maa{\ss}berg, N.~B. Marushchenko, and
  Y.~Turkin.
\newblock Physics in the magnetic configuration space of {W}7-{X}.
\newblock {\em Plasma Physics and Controlled Fusion}, 57(1):014004, November
  2015.

\bibitem{Marushchenko2014}
N.~B. Marushchenko, Y.~Turkin, and H.~Maassberg.
\newblock Ray-tracing code {TRAVIS} for {ECR} heating, {EC} current drive and
  {ECE} diagnostic.
\newblock {\em Computer Physics Communications}, 185(1):165--176, January 2014.

\bibitem{Gusakov2004}
E.~Z. Gusakov and A.~V. Surkov.
\newblock Spatial and wavenumber resolution of doppler reflectometry.
\newblock {\em Plasma Physics and Controlled Fusion}, 46(7):1143--1162, June
  2004.

\bibitem{Jenko2000}
F.~Jenko.
\newblock Massively parallel {Vlasov} simulation of electromagnetic drift-wave
  turbulence.
\newblock {\em Computer Physics Communications}, 125(1-3):196--209, March 2000.

\bibitem{CarraleroISHW2022}
D.~Carralero, T.~Estrada, T.~Windisch, E.~Maragkoudakis, J.~A. Alonso,
  I.~Calvo, J.~M. Garc{\'{i}}a-Regaña, A.~Gonz{\'{a}}lez-Jerez,
  E.~S{\'{a}}nchez, H.~Thienpondt, and J.L. Velasco.
\newblock Recent turbulence investigations in the {TJ}-{II} and {W}7-{X}
  stellarators: experimental characterisation and 3{D} code-based
  interpretation.
\newblock In {\em 23rd International Stellarator-Heliotron Workshop}, 2022.

\bibitem{Told_2013}
D.~Told, F.~Jenko, T.~Görler, F.~J. Casson, and E.~Fable and.
\newblock Characterizing turbulent transport in {ASDEX} upgrade l-mode plasmas
  via nonlinear gyrokinetic simulations.
\newblock {\em Physics of Plasmas}, 20(12):122312, dec 2013.

\bibitem{Biglari1989}
H.~Biglari, P.~H. Diamond, and M.~N. Rosenbluth.
\newblock Toroidal ion-pressure-gradient-driven drift instabilities and
  transport revisited.
\newblock {\em Physics of Fluids B: Plasma Physics}, 1(1):109--118, January
  1989.

\end{thebibliography}

\end{document}